\documentclass[12 pt, article]{article}
\usepackage[left=1in, right=1in, top=0.75in, bottom=0.75in]{geometry}
\usepackage[utf8]{inputenc}
\usepackage{setspace}
\usepackage{graphicx}
\usepackage{authblk}
\usepackage{sectsty}
\usepackage{gensymb}
\usepackage{amsmath}
\usepackage{parskip}
\usepackage{caption}
\usepackage{titlesec}
\usepackage{wrapfig}
\usepackage{multicol}
\usepackage[utf8]{inputenc}
\graphicspath{{images/}}
\sectionfont{\fontsize{13}{16}\selectfont}
\subsectionfont{\fontsize{12}{16}\selectfont}
\subsubsectionfont{\fontsize{11}{16}\selectfont}

\titlespacing{\section}{0pt}{12pt plus 2pt minus 2pt}{12pt plus 2pt minus 2pt}
\usepackage[
  style= chem-acs,
  backend = biber,
  articletitle = true,
  maxbibnames=10,
  minbibnames=10,
]{biblatex}

\renewbibmacro*{name:andothers}{%
  \ifboolexpr{
    test {\ifnumequal{\value{listcount}}{\value{liststop}}}
    and
    test \ifmorenames
  }
    {\addspace\bibstring{andothers}}
    {}}

\title{\Large\textbf {Enantiopurity-Controlled Magnetism in a Two-Dimensional Organic--Inorganic Material}}

\author[1]{P.~Garrett~Hegel}
\author[1,2]{Oscar~Gonzalez}
\author[1]{Mingrui~Li}
\author[1]{Shannon~S.~Fender}
\author[3]{Harishankar~Jayakumar}
\author[3,4]{Archana~Raja}
\author[1]{Ariana~Ray}
\author[1]{Isaac~M.~Craig}
\author[1,4,5*]{D.~Kwabena~Bediako}
\affil[1]{\textit{Department of Chemistry, University of California, Berkeley, CA 94720, USA}}
\affil[2]{\textit{Materials Sciences Division, Lawrence Berkeley National Laboratory, Berkeley, CA 94720, USA}}
\affil[3]{\textit{Molecular Foundry, Lawrence Berkeley National Laboratory, Berkeley, California 94720, USA}}
\affil[4]{\textit{Kavli Energy NanoScience Institute, Berkeley, CA 94720, USA}}
\affil[5]{\textit{Chemical Sciences Division, Lawrence Berkeley National Laboratory, Berkeley, CA 94720, USA}}

\affil[*]{Corresponding author. Email: bediako@berkeley.edu}

\date{April 1, 2026} 

\begin{document}
\maketitle
Extended solids that combine unpaired electron spin and structural chirality can host unconventional magnetic behaviors with potential for electronic technologies. A versatile strategy for creating chiral solids is incorporation of chiral organic molecules into inorganic crystals. However, such hybrid organic–inorganic materials have so far been examined through the lens of absolute chirality, leaving enantiomeric excess (\textit{ee}) underexplored as a tuning parameter. Here, we report two–dimensional (2D) intercalation compounds with controllable \textit{ee} produced by cation exchange of MnPS\textsubscript{3} with chiral organic molecules. We show that these materials' magnetism is determined by intercalant \textit{ee} rather than absolute chirality. Moreover, low–\textit{ee} materials display thermally activated dynamic magnetism absent from enantiopure analogs. These \textit{ee}–dependent magnetic behaviors are explained by local ordering of Mn vacancies, directed by correlated vacancy–intercalant electrostatics and confined molecular packing. Together, these results demonstrate a distinctive tuning strategy for molecule–material hybrids and establish design principles for 2D chiral and magnetically dynamic materials.

\noindent 

\begin{multicols}{2}

\section*{Introduction}

Chirality—the absence of improper symmetry elements (inversion centers, mirror planes, and rotoreflection axes)—can endow solids with properties  absent from otherwise achiral materials, including circularly selective optical responses\supercite{BenMoshe2021, Long2020_2}, asymmetric reactivity\supercite{Soai1999}, spin-polarized charge transport\supercite{Qian2022,Wang2025Topo,Calavalle2022}, and unconventional magnetic textures\supercite{Marty2008, Grigoriev2009}.
Chiral magnetic materials are particularly compelling because they offer opportunities for controlling spin and information flow in next-generation technologies such as spintronics\supercite{Yang2021}, optical spin manipulation\supercite{Wang2025}, and chiral sensing\supercite{Liu2023}. 
Yet, fully realizing these possibilities requires exquisite synthetic control over chirality, including the ability to ensure that a single enantiomorph prevails throughout the material or, alternatively, deterministic control over the extent of chirality. Achieving such long-range control over chirality remains a major challenge: most chiral inorganic crystals form as uncontrolled mixtures of opposite chirality domains\supercite{Togawa2016,Horibe2014,Ohsumi2013,Du2021}, where the inability to control or modulate this heterochirality limits both the deterministic synthesis of chiral materials and the rational tuning of chirality-dependent properties in extended solids.

One route to overcoming uncontrolled heterochirality in inorganic solids is to use homochiral organic molecules to form homochiral organic-inorganic materials. This strategy has been successfully applied in hybrid organic-inorganic perovskites \supercite{Long2020_2}, metal-organic frameworks \supercite{Gong2022, Feng2025}, and two-dimensional (2D) intercalation compounds \supercite{Qian2022,Zhou2006, Tezze2025_2}. In contrast to all-inorganic systems, the degree of homo- and heterochirality in organic components can be precisely controlled by adjusting their enantiomeric excess (\textit{ee}). However, to date, such efforts have focused on preparing enantiopure materials, since it is usually assumed that the strongest chirality-related property responses occur in enantiopure samples\supercite{Panagiotopoulou2025,donckele2014}. This assumption does not always hold true in hybrid solid-state materials, as exemplified by chiral metal-organic perovskites where chiroptical properties such as circular dichroism can depend non-linearly on molecular \textit{ee} with the strongest response occurring at intermediate \textit{ee}\supercite{Panagiotopoulou2025}. Although circular dichroism is a property deeply linked to chirality, results from the hybrid perovskite more broadly suggest the possibility of harnessing molecule-molecule interactions in intermediate-\textit{ee} materials to influence material properties not explicitly related to chirality such as magnetism. Realization of such enantiopurity-controlled properties would then create opportunities for fine property tuning in hybrid materials synthesis by leveraging the continuum of chirality provided by organic molecules with defined \textit{ee}.

The antiferromagnetic van der Waals (vdW) material MnPS\textsubscript{3} can be post-synthetically modified with guest molecules through cation exchange \supercite{Clement1986} as the loss of Mn$^{2+}$ ions is charge-compensated by the intercalation of cationic molecules. The appropriate choice of guest molecule can endow the resulting hybrids with emergent magnetic\supercite{Tezze2025,Evans1995_2} and optical properties\supercite{Coradin1996, Lagadic1997}. 
In this work, we install chirality in otherwise achiral MnPS\textsubscript{3}, space group \textit{C}2/m, through cation exchange with chiral molecules to produce organic--inorganic materials that span a tunable range of \textit{ee} from enantiopure (\textit{R}, \textit{S}) to racemic (\textit{rac}).

Through compositional, structural, spectroscopic, and magnetic characterization, we show that even though the \textit{R}, \textit{S}, and \textit{rac} intercalates share the same long range structures and composition, their magnetic behavior is strongly dictated by the enantiopurity of intercalant molecules. Intercalation of enantiopure molecules turns on ferrimagnetism, whereas \textit{rac} or low-\textit{ee} samples exhibit magnetic behavior that is considerably more compensated (antiferromagnetic). Additionally, low-\textit{ee} materials exhibit thermally activated dynamic magnetism at relatively short timescales (minutes), whereas the enantiopure intercalates remain magnetically invariant at relatively long timescales (months). Together with second harmonic generation (SHG) microscopy, these stark differences in magnetism are understood as arising from the presence/absence of locally ordered vacancy superlattices, which are in turn governed by the disparate packing behavior of enantiopure/racemic molecules confined in the vdW gap. These results demonstrate the importance of—and potential to exploit—chirality and molecular packing in 2D material--molecule hybrid systems with implications for using enantiopurity as a mechanism of synthetic control in chiral 2D materials.

\begin{figure*}[tbh]
    \centerline{\includegraphics[width=150mm]{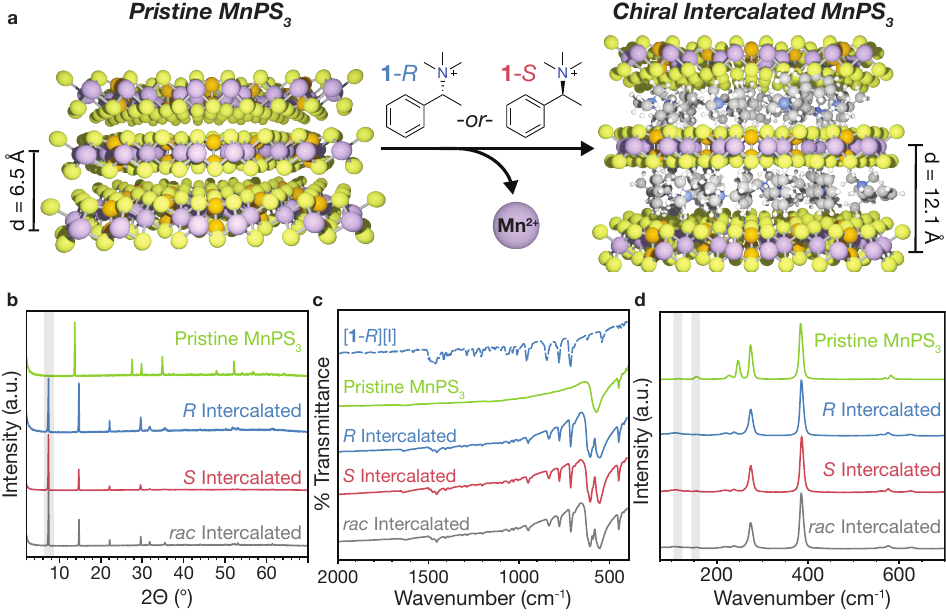}}
    \caption{
     \textbf{(a)} Intercalation of chiral molecules into MnPS\textsubscript{3}. Purple, orange, and yellow spheres represent Mn, P, and S, respectively. \textbf{(b)} Powder X-ray diffraction patterns of pristine and intercalated powders. Data are normalized to the (001) reflection intensity for clarity. The intercalate (001) reflection is highlighted in gray. \textbf{(c)} IR spectra of pristine and intercalated powders. A spectrum of isolated [\textbf{1}-\textit{R}][I] is included for reference. \textbf{(d)} Raman spectra of the pristine and intercalated powders. Modes involving Mn are highlighted in gray. 
    }\label{fig:M1}
\end{figure*}

\section*{Results and Discussion}

\subsection*{Chemical and Structural Characterization}
Cation exchange-driven intercalation of MnPS\textsubscript{3} proceeds via Mn\textsuperscript{2+} vacancy generation and cation uptake according to\supercite{Bernier1993}:
\begin{center}
    MnPS\textsubscript{3} + (Guest)\textsuperscript{+}\textsubscript{(aq)}  $\longrightarrow$ Mn$_{1-x}$PS\textsubscript{3}(Guest)$_{2x}$(H\textsubscript{2}O)$_{y}$ + $x$Mn\textsuperscript{2+}\textsubscript{(aq)}
\end{center}
Direct intercalation of MnPS\textsubscript{3} occurs readily with small, minimally charged intercalants\supercite{Clement1979,Iton2024,Clement1986}.
Larger cations can be installed via a two-step process\supercite{LaCroix1993,Coradin1996,Clement1981,Leaustic1995}, but the preintercalation step can compromise the crystallinity of the final material and is therefore undesirable\supercite{Zhang2007, Zhou2006}.
To achieve one-step intercalation with a chiral molecule, we synthesized the monovalent cation \textit{N,N,N}-trimethylmethylbenzylammonium iodide in its enantiopure and racemic forms ([\textbf{1}-\textit{R/S/rac}][I]) (Figure 1a). Circular dichroism spectra of intercalant molecule solutions (Figure S1) establish their chirality.

Intercalation in MnPS\textsubscript{3} increases the material's interlayer spacing as intercalant cations and any accompanying solvent molecules force expansion of the interlamellar region (Figure 1a)\supercite{Clement1986,Iton2024}.  Reaction progress is monitored through powder X-ray diffraction (PXRD) as the pristine (001) reflection disappears and is replaced with a new (001) reflection corresponding to an interlayer spacing increase from 6.5 Å to 12.1 Å (Figure 1b)\supercite{Evans1995,Tezze2025}. As a result of the material's preferred orientation and low in-plane crystallinity, space group indexing for the intercalate was unsuccessful even on higher-resolution synchrotron diffraction patterns (Figure S2). Monitoring intercalation of bulk MnPS\textsubscript{3} flakes reveals heat is necessary to drive the reaction to completion (Figure S3). Thermal activation implies a kinetic barrier to intercalation, and this kinetic barrier is responsible for small amounts of unintercalated phase seen in high-resolution PXRD patterns. In keeping with previous reports\supercite{Evans1995, Iton2024}, the resulting crystals are insufficiently ordered to allow for structural determination by single crystal X-ray diffraction.

Infrared (IR) spectroscopy confirms intercalation; vibrations corresponding to \textbf{1}, which are absent in the pristine MnPS\textsubscript{3}, are present in the intercalates (Figure 1c). 
Raman spectra of all samples are dominated by modes associated with the ethane-like P\textsubscript{2}S\textsubscript{6}\textsuperscript{4-} polyanion (Figure 1d)\supercite{Mathey1980, Sun2019, Grasso1989, Clement1986}. These modes shift by less than 10 cm\textsuperscript{-1} upon intercalation (SI Table 1), indicating that bonding within the polyanion changes minimally upon intercalation. The intensity of modes associated with Mn--S bonds attenuate significantly with intercalation\supercite{Joy1992,Tezze2025}, consistent with the generation of Mn vacancies due to cation exchange. Select high-energy modes originating from the organic molecules are weakly observable in the intercalates (Figure S4).

Mn and P content of the samples are determined by inductively coupled plasma optical emission spectroscopy (ICP-OES) analysis (Figure S5) and used to evaluate organic intercalant loading (see Methods). From this analysis, we calculate nominal compositions of the intercalates presented in Figure 2a to be: Mn\textsubscript{0.84}(\textbf{1}-\textit{R})\textsubscript{0.31}P\textsubscript{0.96}S\textsubscript{3}, Mn\textsubscript{0.83}(\textbf{1}-\textit{S})\textsubscript{0.33}P\textsubscript{0.96}S\textsubscript{3}, and Mn\textsubscript{0.83}(\textbf{1}-\textit{rac})\textsubscript{0.34}P\textsubscript{0.96}S\textsubscript{3}. Compositions vary slightly from batch to batch but settle around 1/6 Mn replacement (Mn$_{1-x}$ = Mn\textsubscript{0.83}) as observed in many other in many other organic and organometallic intercalated systems (SI Note 1)\supercite{Evans1995_2,Clement1981,Lagadic1997,Floquet2002,Zhou2006,Tezze2025}. The small batch-to-batch compositional variation may stem from small amounts of residual pristine phase and from compositional variation inherent to the disordered nature of MnPS\textsubscript{3} intercalation \supercite{Iton2024, Tezze2025}. Thermogravimetric analysis (TGA) estimates the amount of cointercalated water as less than 1 molecule of water per guest cation (Figure S6). Compositional analyses were additionally supported by CHNS analysis (SI Note 1).

\subsection*{Magnetic Properties}

\subsubsection*{Static Magnetism}

\begin{figure*}[tbh]
    \centerline{\includegraphics[width=82.5mm]{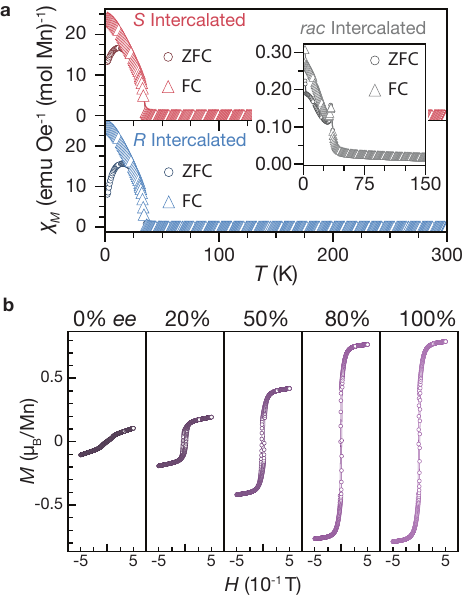}}
    \caption{
     \textbf{(a)} Magnetic susceptibility as a function of temperature of the \textit{R}, \textit{S}, and \textit{rac} intercalated samples measured at 100 Oe. \textbf{(b)} Representative magnetic hysteresis curves at 2 K for varying \textit{ee} intercalates.
    }\label{fig:M2}
\end{figure*}

Pristine MnPS\textsubscript{3} is an antiferromagnet ($T_\textrm{N}$ = 78 K) where each high-spin Mn\textsuperscript{2+} site is antiferromagnetically coupled to its nearest neighbors via superexchange (Figure S7)\supercite{Taylor1973,Kurosawa1983,Okuda1986}. Intercalation of \textbf{1}-\textit{R} and \textbf{1}-\textit{S} into MnPS\textsubscript{3} powders and resultant loss of about 1/6 Mn$^{2+}$ ions leads to uncompensated magnetic moments consistent with ferrimagnetism \supercite{Evans1995_2, Tezze2025} with a $T_\textrm{C}$  of 39 K and a bifurcation between zero-field-cooled (ZFC) and field-cooled (FC) $\chi_{M}$ vs. $T$ plots (Figure 2a). At high temperature, Curie--Weiss analysis of \textit{R} and \textit{S} intercalates are consistent with predominantly antiferromagnetic interactions with Curie--Weiss temperatures ($\theta_\textrm{{CW}}$) of $-95$ K and $-94$ K, respectively (Figure S8). In addition, these fits yield effective magnetic moments, $\mu_{\mathrm{eff}}$, of 6.26 and 6.12 $\mu_{B}$/Mn for \textit{R} and \textit{S} respectively, slightly higher than the calculated spin-only moment of 5.92 $\mu_{B}$/Mn for $S=5/2$ Mn\textsuperscript{2+}. A collection of aligned \textit{S}-intercalated bulk crystals display a greater out-of-plane magnetization ($M$) than in-plane $M$ (Figure S9), indicative of an out-of-plane easy axis, as reported for other ferrimagnetic MnPS\textsubscript{3} intercalates\supercite{Evans1995, Tezze2025}.

Consistent with ferrimagnetism, measurements of magnetization, $M$, as a function of magnetic field, $H$, yields S–shaped traces for both \textit{R} and \textit{S} intercalates with soft hysteresis (Figure S10). At 2 K, the \textit{R} and \textit{S} samples show coercivities, $H_c$, of 105 Oe and 154 Oe, respectively. 
Though no additional features are observed upon measuring to 12 T, the magnetization does not saturate at that field (Figure S11). Hysteresis loops close above 30 K, and the S–shaped curve disappears entirely above 40 K. Ferrimagnetism in enantiopure samples is qualitatively reproducible between batches (Figure S12).  We attribute slight differences in magnetization metrics ($H_c, \mu_{\mathrm{eff}}, \theta_\textrm{{CW}}$) between batches and between \textit{R} and \textit{S} enantiomers to slight sample-to-sample differences in Mn vacancy order/disorder.

The magnetic behavior of the freshly prepared \textit{rac} intercalate is substantively different from that of the enantiopure intercalates. At high temperatures, the \textit{rac} intercalate maintains antiferromagnetic correlations with $\theta_\textrm{{CW}}$ of $-100$ K and $\mu_{\mathrm{eff}}$ of 5.98 $\mu_{B}$/Mn, values comparable to those of the enantiopure materials. However, at low temperature, the racemic compound displays a susceptibility two orders of magnitude lower than those of the enantiopure intercalates (Figure 2a). Although lower in magnitude, the \textit{rac} sample retains a sharp susceptibility upturn at 38 K as well as a slight bifurcation between ZFC- and FC-curves, which are consistent with some ferrimagnetic character and small magnetic coercivity.
A 2 K $M$($H$) trace of the racemic compound displays double–lobed hysteresis with a maximum $H_c$ of 69 Oe and lower overall moment than that of the enantiopure samples (Figure S10). These magnetization behaviors--namely, a considerably lower moment than in the enantiopure compounds and a double-lobed hysteresis--are qualitatively reproducible across multiple racemic samples  (Figure S13). Additionally, beyond the extremes of enantiopure and racemic intercalates, the amount of uncompensated magnetic moment is tunable through modulation of the \textit{ee} of the intercalated molecules, with low–\textit{ee} materials displaying a lower moment than high–\textit{ee} materials (Figure 2b, S14). These differences in magnetic behavior among samples with disparate \textit{ee} are not explained by compositional differences (Figure S15), suggesting a distinctive mechanism underpinned by enantiopurity.

\subsubsection*{Dynamic Magnetism in Racemic Intercalates}

\begin{figure*}[tbh]
	\centerline{\includegraphics[width=82.5mm]{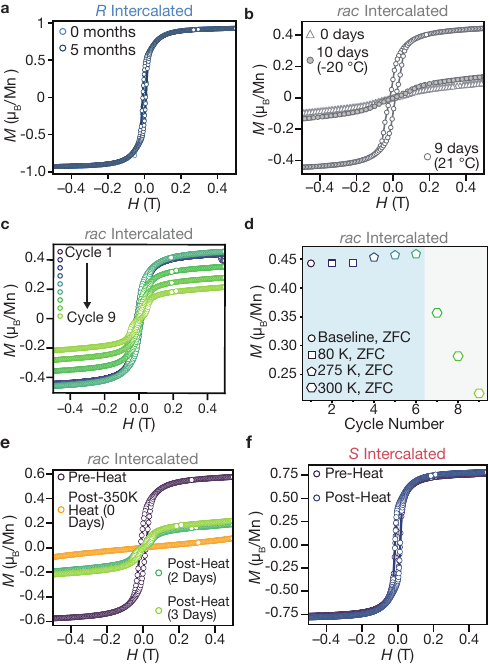}}
	\caption{ \textbf{(a-b)} Enantiopure and racemic intercalate magnetization as a function of applied field measured after aging under different storage conditions. \textbf{(c)} Racemic intercalate magnetization after thermal cycling. \textbf{(d)} Maximum magnetization at 0.5 T from \textbf{(c)} as a function of cycle number and cycling conditions. Blue shading denotes thermal cycles conducted under 300 K. \textbf{(e-f)} Racemic and enantiopure intercalate magnetization before and after a 12 hour 350 K heating and subsequent room temperature aging. All measurements are conducted at 2 K. }
	\label{fig:M3}
\end{figure*}

In enantiopure intercalates, $M$($H$) curves measured 5 months apart are indistinguishable after aging under ambient conditions (Figure 3a). In contrast, the racemic material's uncompensated moment changes dramatically upon aging over approximately one week (Figure 3b). The evolution of this magnetic behavior is temperature dependent; samples stored at room temperature evolve more quickly than samples stored at $-20$ °C. Aging behavior was also observed in samples with \textit{ee} $\leq 50 \%$.

The magnetization of fresh racemic samples evolves with thermal cycling (Figure 3c-d). The sample was cooled to 2 K for an initial $M$($H$) sweep, then warmed to a specific temperature for aging over 30 minutes, and finally cooled back to 2 K for another $M$($H$) measurement. The process was then repeated for other aging temperatures. Magnetization changes negligibly upon cycling to 80 K and 275 K, but substantial changes are observed upon cycling to 300 K. We emphasize that the sample does not leave the PPMS chamber during this process, and no changes in material structure are detectable by PXRD after this thermal cycling (Figure S16). Cycle-dependent magnetization behavior is observed across multiple fresh racemic samples, although the changes in net uncompensated moment (increase or decrease) vary from sample to sample. For the samples shown in Figure 3d and Figure S17b, cycling results in a decrease in net magnetization. In contrast, as shown in Figure S17a, some samples display an increase in net magnetization with cycling. Moreover, samples do not converge to a singular magnetization value (Figure S17c,d) by cycling. In aged racemic samples, which possess a relatively high saturation magnetization, ferrimagnetic behavior can be almost fully quenched by heating the sample at 350 K for 12 hours (Figure 3e). Uncompensated magnetic moments and ferrimagnetic hysteresis then return after aging at room temperature over 2--3 days. In contrast, enantiopure samples show negligible changes with heating to 350 K (Figure 3f).

\subsubsection*{Origins of Enantiopurity–Based Magnetism}

Ferrimagnetism in intercalated MnPS\textsubscript{3} arises from intralayer ordering of Mn vacancies\supercite{Evans1995,Evans1995_2}. In pristine MnPS\textsubscript{3}, the magnetic moments on Mn\textsuperscript{2+} ions are fully compensated as a result of equal and oppositely aligned spin sublattices (Figure 4a,c)\supercite{Kurosawa1983}. Intercalation disrupts this balance by removing Mn\textsuperscript{2+} ions, leading to incomplete compensation that depends sensitively on the spatial arrangement of the induced Mn vacancies\supercite{Tezze2025}. A 1/6 removal of Mn\textsuperscript{2+} ions confined to a single spin sublattice results in ferrimagnetism, producing one uncompensated $S=5/2$ site for every five Mn ions ($M_{\text{saturation}}$ = 1 $\mu_B$ per Mn; see SI Note 2)\supercite{Evans1995_2}. By contrast, a 1/6 random removal of Mn\textsuperscript{2+} ions—such that vacancies are statistically distributed between both spin sublattices—preserves complete spin compensation and results in antiferromagnetic order ($M=0$). Experimentally, many MnPS\textsubscript{3} intercalates exhibit ferrimagnetic behavior with a saturation magnetization of approximately 1 $\mu_B$ per Mn\supercite{Evans1995_2}.

\begin{figure*}[tbh]
	\centerline{\includegraphics[width=160mm]{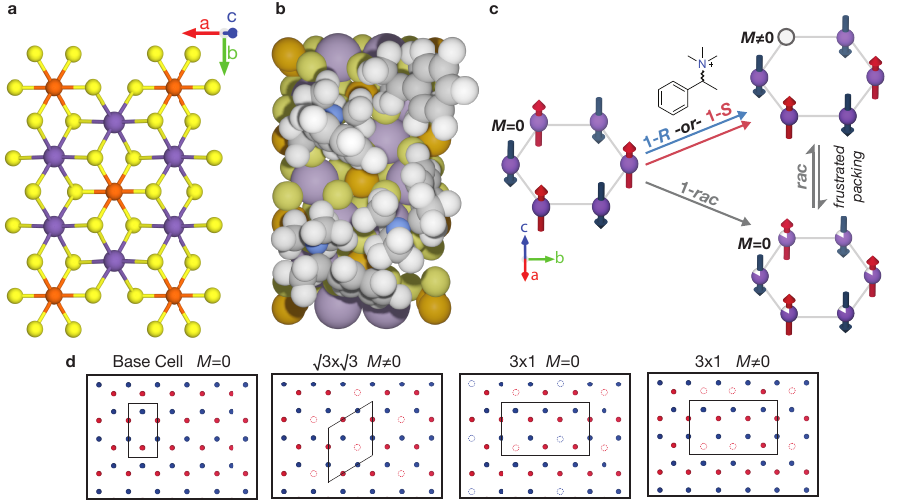}}
	\caption{ \textbf{(a)} The structure of hexagonally arranged Mn atoms within a single MnPS\textsubscript{3} layer. \textbf{(b)} Top down spacefill model of intercalated molecules on a slab of MnPS\textsubscript{3} with water molecules omitted. \textbf{(c)} Limiting cases of ferrimagnetically ordered Mn vacancies and evenly distributed, antiferromagnetic Mn vacancies after intercalation-induced removal of 1/6 Mn sites. \textbf{(d)} Possible ordered superlattices of Mn vacancies. Different net magnetizations are possible depending on the combination of  superlattice cell and vacancy site(s).}
	\label{fig:M4}
\end{figure*}

Here, we find that chiral intercalated systems appear to span between these two extremes, with the \textit{ee} dictating the preponderance of ferrimagnetic or antiferromagnetic domains.  Enantiopure samples display magnetization values ranging from 0.75 to 0.93 $\mu_{B}$ per Mn at 2 K and 0.5 T. These values are consistent with a majority of ferrimagnetic domains interspersed with a small minority of antiferromagnetic regions. Freshly prepared racemic samples display magnetization values ranging from 0.07 to  0.19 $\mu_{B}$ per Mn at 2 K and 0.5 T, suggesting these samples are comprised of a large majority of antiferromagnetic domains. A mixture of interacting antiferromagnetic and ferrimagnetic domains can also explain the observed double-lobed hysteresis\supercite{Bennett2005,Andrez2017}.

These observations that magnetism of intercalated MnPS\textsubscript{3} (dictated by vacancy ordering) depends on the enantiopurity of chiral intercalant cations can be understood by drawing an analogy between the distinct packing motifs and morphologies reported for enantiopure and racemic molecular thin films\supercite{Chen2019,Liu2013} and the packing of chiral molecules in the interlamellar region. Because Mn vacancies and intercalant cations are electrostatically attracted, the intercalant configuration influences vacancy ordering and, consequently, the net magnetization of the material\supercite{Tezze2025}. Notably, molecular identity alone cannot explain the magnetic differences, as racemic and enantiopure intercalates contain the same molecules. At a 1/6 replacement of Mn ions, intercalants are densely confined within the interlamellar region (Figure 4b), where their packing becomes a key determinant of magnetic order. We hypothesize that identical enantiomers more readily adopt an ordered arrangement under this confinement than do mixtures of opposite enantiomers (Figure 4c). In low-\textit{ee} intercalates, steric incompatibilities among intercalants, combined with competing Coulombic interactions between intercalants and Mn vacancies, would frustrate molecular packing\supercite{Chen2019,Liu2013}. This frustration should in turn disrupt Mn-vacancy order, suppressing ferrimagnetism and producing a system that is magnetically compensated over the length scale of a few unit cells.

The dynamic magnetism observed in the racemic material can also be rationalized by frustrated chiral molecular packing. Although we attribute ferri- and antiferromagnetic behavior to domains of periodically ordered Mn vacancies, the precise vacancy arrangement within these domains is not uniquely defined; multiple vacancy superlattices can support either magnetic state (Figure 4d). Moreover, the known lability of Mn\textsuperscript{2+} in MnPS\textsubscript{3}, which enables cation exchange and intercalation at and above room temperature, implies that Mn vacancies remain mobile on experimental timescales. Under these conditions, elevated temperatures allow the system to sample competing vacancy configurations, as well as intermediate or mixed arrangements (Figure 4c). Cooling the racemic intercalate for measurement arrests vacancy motion, capturing a snapshot of the instantaneous ensemble of vacancy configurations. This thermally activated vacancy mobility accounts for the pronounced evolution of magnetic behavior observed in racemic intercalates when aged at high temperatures. Consistently, heating to 350 K disrupts vacancy order in the racemic material, quenching ferrimagnetism, which is then subsequently recovered during room temperature aging as the system gradually moves toward a more ordered state. By contrast, the enantiopure material exhibits little sensitivity to heating at 350 K, suggesting either kinetic trapping in a ferrimagnetic configuration or raid re-establishment of ferrimagnetically ordered vacancies following thermal disordering.

\begin{figure*}[tbh]
	\centerline{\includegraphics[width=70mm]{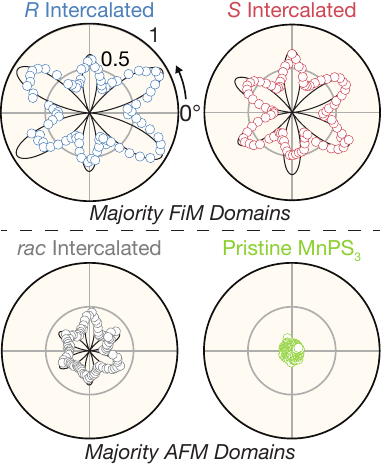}}
	\caption{ Polarization dependence of SHG collected at room temperature from pristine and intercalated bulk crystals. Solid black lines are fits to the data.
}
	\label{fig:M5}
\end{figure*}

\subsection*{Second Harmonic Generation}

To probe the proposed Mn-vacancy ordering, we performed confocal second-harmonic generation (SHG) microscopy on intercalated MnPS\textsubscript{3}. At room temperature, the centrosymmetric structure of pristine MnPS\textsubscript{3} suppresses SHG via the dominant electric-dipole (ED) mechanism, allowing only the much weaker electric-quadrupole (EQ) contribution\supercite{Ni2021}. As a result, SHG from pristine MnPS\textsubscript{3} is often weak or undetectable\supercite{Ni2021,Chu2020}, and in our measurements lies below the detection limit (Figure 5). In contrast, clear SHG signals are observed from \textit{R}, \textit{S}, and \textit{rac} intercalated MnPS\textsubscript{3} crystals, with the strongest response arising from enantiopure intercalates. The intercalated samples exhibit sixfold or pseudo-sixfold polarization dependences, with modest spot-to-spot variations consistent with previous reports on pristine MnPS\textsubscript{3}\supercite{Chu2020}. Notably, MnPS\textsubscript{3} intercalated with the centrosymmetric cation NMe\textsubscript{4}\textsuperscript{+} produces a stronger SHG signal than the enantiopure chiral intercalates (Figure S18), indicating that the enhanced SHG originates primarily from the inorganic lattice rather than from low-symmetry molecules within the vdW interface.

The enhanced SHG observed upon intercalation is consistent with symmetry breaking induced by Mn-vacancy ordering. Assuming preservation of the host lattice symmetry and monoclinic stacking, the vacancy superlattices illustrated in Figure 3d correspond to the space groups \textit{C}2/m, \textit{C}2, \textit{P}2/c, and \textit{P}2\textsubscript{1}. The antiferromagnetic (AFM) configurations retain inversion symmetry (\textit{C}2/m and \textit{P}2/c), whereas the ferrimagnetic (FiM) configurations break inversion symmetry (\textit{C}2 and \textit{P}2\textsubscript{1}). Such noncentrosymmetric domains permit ED-allowed SHG and therefore provide a natural explanation for the enhanced SHG signal in ferrimagnetic samples. This interpretation is supported by the correlation between SHG intensity and magnetic moments measured by ensemble magnetometry, consistent with varying fractions of ferrimagnetic vacancy-ordered domains. As an alternative contribution, intercalation may also enhance EQ SHG by increasing the polarizability of the lattice through the introduction of additional species into the vdW interface. However, under normal-incidence geometry, the polarization dependences of EQ SHG from pristine MnPS\textsubscript{3} and ED SHG arising from vacancy-induced symmetry breaking share the same functional form and are therefore indistinguishable based on polarization analysis alone (SI Note 3). Taken together, the agreement between spatially resolved SHG and bulk magnetometry supports a picture of heterogeneous Mn-vacancy superlattice domains.

\section*{Conclusions}

We show that the enantiopurity of organic molecules intercalated into MnPS\textsubscript{3} provides a powerful handle for controlling magnetism in a vdW solid. Integrated structural, spectroscopic, and magnetic measurements suggest that the dynamic magnetism observed in low-\textit{ee} intercalates originates from frustrated chiral molecular packing coupled to Coulombic interactions between intercalants and Mn\textsuperscript{2+} vacancies. More broadly, our results establish molecular chirality and packing frustration as design parameters for programming collective, reconfigurable behavior in organic-inorganic hybrid materials. Extending this approach to other low-symmetry intercalants and layered hosts may enable the deliberate creation of dynamically adaptive magnetic, electronic, or optical states in vdW solids. Finally, the pronounced differences observed here between racemic and enantiopure intercalates highlight an important caution: racemic intercalated materials should not be assumed to serve as straightforward `achiral controls' when benchmarking chiral versus achiral systems. Instead, careful structural validation is required to confirm that racemic and enantiopure intercalates are truly isostructural.

\section*{Methods}

\subsection*{Material Preparation}

\subsubsection*{Starting Materials}
Commercial chemicals, Mn powder (Beantown Chemical, 99.95\%), S (Acros, 99.999\%), red P (99.99\%), P\textsubscript{2}S\textsubscript{5} (Sigma, 99\%), \textit{R/S/rac}-$\alpha$-methylbenzylamine (Sigma, 98\%), K\textsubscript{2}CO\textsubscript{3} (Sigma $\geq$99\%), MeI (Sigma, 99\%), NMe\textsubscript{4}Cl (Sigma, $\geq$98\%) DCM (Fisher, HPLC grade), MeOH (Fisher, ACS grade), were used as received without further purification. MeCN (Fisher, HPLC grade) was dried on a solvent purification system prior to use.

\subsubsection*{Intercalant Synthesis}

[\textbf{1}-\textit{R/S/rac}][I] were synthesized via procedures adapted from the literature\supercite{Zuo2018, LaCour2002}. \textit{R/S/rac}-$\alpha$-methylbenzylamine (0.45 mL, 4 mmol), K\textsubscript{2}CO\textsubscript{3} (1.38 g, 10 mmol), and dry MeCN (15 mL) were combined in an oven dried 50 mL flask under nitrogen atmosphere. MeI (1.25 mL, 20 mmol) was then added dropwise to the reaction to the reaction while stirring. The reaction was then stirred under nitrogen for 18 hours at room temperature, evaporated to dryness, and then resuspended in 50 mL of DCM. The mixture was filtered and the solvent evaporated to yield \textbf{1}-\textit{R/S/rac} as a white solid. Yield 1.054 g, 90 \%: $^1H$ NMR, (500 MHz, CDCl\textsubscript{3}): $\delta$ 7.64 (m, 2H), 7.50-7.45 (m, 3H), 5.40 (q, 1H), 3.39 (s, 9H), 1.86 (d, 3H).  $^{13}$C NMR, (500 MHz, CDCl\textsubscript{3}): $\delta$ 132.4, 130.9, 129.3, 77.3, 51.6, 15.5 HRMS (ESI+) \textit{m/z} calcd for \textbf{1}: 164.144, Found  \textit{R}: 164.141, \textit{S}: 164.146, \textit{rac}: 164.147

\subsubsection*{Host Lattice Synthesis}
Powders of MnPS\textsubscript{3} were synthesized by a P\textsubscript{2}S\textsubscript{5} self flux method adapted from the literature\supercite{Iton2024,Chica2021}. Mn powder (132 mg, 2.4 mmol), S (39 mg, 1.2 mmol), and P\textsubscript{2}S\textsubscript{5} (293 mg, 1.32 mmol) were ground to a homogenous mixture in a mortar and pestle in an Ar glovebox. The powder was then transferred to a 1.4 cm inner-diameter silica tube before being sealed at 20 cm length under vacuum ($<$ 30 mTorr). The sample was then heated to 650 °C at 1 °C/min and held at that temperature for 72 hours. After cooling to room temperature the ampoule was positioned such that the sample remained in the furnace and a portion of the tube sat outside the furnace. The furnace was heated to 300 °C and dwelled for 24 hours to remove excess P\textsubscript{2}S\textsubscript{5}. The green product (a mixture of powder and larger crystals) was collected, ground to a powder in a mortar and pestle, and stored in air.

Bulk crystals of MnPS\textsubscript{3} were synthesized through chemical vapor transport. Mn powder, S, and red P in a 1:3:1 molar ratio (131 mg total) was ground to a homogeneous mixture in a mortar pestle in an Ar filled glovebox. The powders were sealed as above at a 35 cm length. The ampoules were heated at 1 °C/min to a 675-600 °C gradient in a two zone furnace with the unreacted starting material at the hot end. After holding the gradient for 2 weeks and cooling to room temperature, green plate-like crystals were collected from the center of the ampoule.
\subsubsection*{MnPS\textsubscript{3} Intercalation}
In a typical powder intercalation, a one dram vial was charged with a stir bar and 100 mg of MnPS\textsubscript{3}. 2.5 mL of 100 mM [\textbf{1}][I] solution prepared in Milli-Q water was then added to the vial. The reaction was heated and stirred at 65 °C for 4 days before replacing the solution and heating for another 4 days. The reaction mixture was filtered and washed with water followed by MeOH. The light green powder was then dried under vacuum overnight. 

Intercalation of bulk crystals was similar, except the crystals were not stirred and significantly less mass of MnPS\textsubscript{3} was used. Bulk crystals also appeared light green after intercalation.

\subsection*{Materials Characterization}

\subsubsection*{Intercalant Characterization}
Circular dichroism spectra were collected on a Jasco J-815 CD Spectropolarimeter. Samples were prepared as 3 mM solutions in ethanol. Data was averaged over 3 accumulations with a speed of 50 nm/min. $^1$H and $^{13}$C NMR were conducted on a Bruker NEO-500. Chemical shifts were referenced to internally referenced to residual solvent (CDCl\textsubscript{3} ($\delta$ ($^1$H) = 7.26 ppm and $\delta$ ($^{13}$C) = 77.2 ppm).  HRMS data was collected on a Perkin Elmer Axion 2 TOF using ESI in positive mode.

\subsubsection*{PXRD}
All PXRD data was collected using a  Bruker D8-Advance diffractometer using Cu K$\alpha$ ($\lambda$ = 1.5406 Å) radiation in Bragg-Bretano geometry unless otherwise indicated. Synchrotron PXRD data was collected at the Stanford Synchrotron Radiation Lightsource beamline 2-1 via an automated diffraction setup.\supercite{Stone2023} Samples were run in Kapton capillaries using $\lambda$ = 0.729659 Å radiation.

\subsubsection*{Compositional Analysis}
ICP-OES analysis was conducted on a Perkin Elmer ICP Optima 7000 DV Spectrometer. To prepare samples, approximately 6 mg of sample was digested in 1 mL of 70\% nitric acid at 80 °C for 4 hours. The digested sample was then diluted twice to reach 2\% nitric acid concentration before being run on the ICP instrument. By calculating the ratio of Mn:P and normalizing the pristine Mn content to 1, the concentration of Mn vacancies generated by intercalation can be determined. Since Mn\textsuperscript{2+} vacancies are charge balanced by two monovalent cations, the concentration of intercalants was accordingly deduced from ICP. CHNS analysis was conducted by the UC Berkeley Microanalytical Facility using a PerkinElmer 2400 Series II combustion analyzer.

TGA was conducted on a TA instruments Discovery TGA 5500. Sample (4.5 - 6.5 mg) was added evenly to an aluminum pan and heated at 5 °C/min from room temperature (20-24 °C) to 200 °C under nitrogen flow (25 mL/min). Water content was calculated by recording mass loss at 150 °C and assuming that mass loss to be entirely water.

\subsubsection*{Spectroscopic Measurements}

FTIR spectra were collected from sample pressed into KBr pellets on a Thermo Fisher Scientific Nicolet iS20 FTIR in transmission mode.

Conforcal raman spectroscopy was conducted using a 532 nm laser on a Horiba Multiline LabRam Evolution. Laser power was approximately 35 $\mu$W, and data was collected using a 600 gr/mm grating.

Confocal SHG data were collected on a custom built laser table setup at the Molecular Foundry Laser Lab at Lawrence Berkeley National Lab (LBNL).
The data was collected at normal incidence to the crystal using an incident 910 nm laser in parallel polarization geometry with polarization angle set by a half waveplate (Figure S16). Since crystallographic axes of the samples were unknown upon mounting, for ease of comparison polarization data was rotated during post-processing such that angles of maximum polarization appear coincident across samples.

\section*{Supporting Information}
Expanded compositional analysis; calculating saturation magnetization as a function of composition; SHG polarization dependence; detailed intercalate Raman; CD of intercalant molecules; additional diffraction data from powders and bulk flakes; TGA and DTG of intercalates; additional magnetometry data; SHG data from an achiral intercalate; SHG setup; \textsuperscript{1}H NMR and \textsuperscript{13}C NMR spectra for intercalant molecules (PDF)

\section*{Acknowledgements}
This material is based upon work supported by the US National Science Foundation Quantum Sensing Challenges for Transformational Advances in Quantum Systems (QuSeC-TAQS) program, under award no. 2326838. We thank Adrian J. Huang for assistance with TGA measurments. NMR spectra were collected at the College of Chemistry Pines Magnetic Resonance Center’s Core NMR facility at the University of California, Berkeley. Work at the Molecular Foundry was supported by the Office of Science, Office of Basic Energy Sciences, of the U.S. Department of Energy under Contract No. DE-AC02-05CH11231. The  Lawrence Berkeley National Laboratory (LBNL) Catalysis Laboratory provided HR-ESI-MS and CD instrumentation. Use of the Stanford Synchrotron Radiation Lightsource, SLAC National Accelerator Laboratory, is supported by the US Department of Energy, Office of Science, Office of Basic Energy Sciences under contract No. DE-AC02-76SF00515. DKB also acknowledges support from the Philomathia Foundation and the Heising-Simons Faculty Fellowship.

\printbibliography

\end{multicols}

\end{document}


\maketitle

\tableofcontents

\doublespacing
\clearpage

\section{Supplementary Note 1: Compositional analysis}
Based on ICP analysis compositions for samples are:\\
\textit{R} Intercalated (Batch 1): Mn\textsubscript{0.84}(\textbf{1}-\textit{R})\textsubscript{0.31}P\textsubscript{0.96}S\textsubscript{3} \\
\textit{R} Intercalated (Batch 2): Mn\textsubscript{0.79}(\textbf{1}-\textit{R})\textsubscript{0.42}P\textsubscript{0.95}S\textsubscript{3} \\
\\
\textit{S} Intercalated (Batch 1): Mn\textsubscript{0.83}(\textbf{1}-\textit{S})\textsubscript{0.33}P\textsubscript{0.96}S\textsubscript{3} \\
\textit{S} Intercalated (Batch 2): Mn\textsubscript{0.79}(\textbf{1}-\textit{S})\textsubscript{0.42}P\textsubscript{0.95}S\textsubscript{3} \\
\\
\textit{rac} Intercalated (Batch 1): Mn\textsubscript{0.83}(\textbf{1}-\textit{rac})\textsubscript{0.34}P\textsubscript{0.96}S\textsubscript{3} \\
\textit{rac} Intercalated (Batch 2): Mn\textsubscript{0.80}(\textbf{1}-\textit{rac})\textsubscript{0.40}P\textsubscript{0.95}S\textsubscript{3} \\
\textit{rac} Intercalated (Batch 3): Mn\textsubscript{0.83}(\textbf{1}-\textit{rac})\textsubscript{0.34}P\textsubscript{0.95}S\textsubscript{3} \\
\textit{rac} Intercalated (Batch 4): Mn\textsubscript{0.84}(\textbf{1}-\textit{rac})\textsubscript{0.32}P\textsubscript{0.95}S\textsubscript{3} \\
\\
\\
20 \% ee Intercalated: Mn\textsubscript{0.87}(\textbf{1}-\textit{S/R})\textsubscript{0.27}P\textsubscript{0.99}S\textsubscript{3} \\
50 \% ee Intercalated: Mn\textsubscript{0.84}(\textbf{1}-\textit{S/R})\textsubscript{0.33}P\textsubscript{0.99}S\textsubscript{3} \\
80 \% ee Intercalated: Mn\textsubscript{0.85}(\textbf{1}-\textit{S/R})\textsubscript{0.29}P\textsubscript{1.02}S\textsubscript{3} \\

Full compositions for select batches were calculated by integrating results from CHNS analysis, ICP, and TGA. Considering the existence of partially intercalated phases, variable solvation, the number of elements involved, and inherent instrumental error, multiple similar compositions can equivalently fit the experimental results. Compositions presented in the main text are calculated via ICP alone to allow for the most straightforward comparisons. Combined ICP and CHNS analysis results of representative batches are presented here for completeness:\\
\\
\textit{R} Intercalated (Batch 1): Mn\textsubscript{0.84}(\textbf{1}-\textit{R})\textsubscript{0.32}P\textsubscript{0.96}S\textsubscript{3.1}(H\textsubscript{2}O)\textsubscript{0.5} 
Found (Calculated): Mn 18.57 (19.41), P 11.94 (12.53), C 18.12 (18.00), N 1.91 (1.89), H 2.37 (2.90), S 43.96 (41.89) \\
\\
\textit{S} Intercalated (Batch 1): Mn\textsubscript{0.83}(\textbf{1}-\textit{S})\textsubscript{0.33}P\textsubscript{0.96}S\textsubscript{3.1}(H\textsubscript{2}O)\textsubscript{0.1} 
Found (Calculated): Mn 19.87 (19.77), P 12.94 (12.84), C 18.51 (19.03), N 1.88 (2.02), H 2.31 (2.70), S 44.37 (42.94)\\
\\
\textit{rac} Intercalated (Batch 1): Mn\textsubscript{0.83}(\textbf{1}-\textit{rac})\textsubscript{0.17}P\textsubscript{0.96}S\textsubscript{3.1}(H\textsubscript{2}O)\textsubscript{0.1}
Found (Calculated): Mn 20.22 (19.56), P 13.24 (12.78), C 18.20 (19.45), N 1.88 (1.96), H 2.28 (2.76), S 44.58 (42.71)\\

\section{Supplemantry Note 2: Modeling saturation magnetization as a function of composition}

The saturation magnetization of MnPS\textsubscript{3} intercalates can be modeled as a function of Mn replacement. In a pristine MnPS\textsubscript{3} sample, there is $1$ Mn per formula unit. Accordingly, there is $1/2$ spin-up Mn and $1/2$ spin-down Mn per formula unit. Assume all removed Mn atoms belong to the same spin-up sublattice. As the amount of removed Mn sites, $x$, varies between $x = 0$ and $x = 1/2$, the amount of spin-up Mn remaining is $1/2-x$, while the amount of spin-down Mn remains $1/2$. Accordingly, the amount of uncompensated spin-down sites per formula unit is given as $1/2-(1/2-x) = x$. Further, there are $1-x$ remaining Mn per formula unit.

Assume only uncompensated spin sites generated by intercalation contribute to net magnetization in intercalated MnPS\textsubscript{3}. At low temperatures, saturation magnetization, $M$, for an individual spin site is given by $g\mu_{\beta}S$. Therefore, $M(x)$ as measured in $\mu_{\beta}$/Mn where  $0\leq x \leq 1/2$ is described by $M(x) = \frac{xgS}{1-x}$ since each of the $x$ uncompensated spin-down sites contributes $Sg$ moment and $1-x$ normalizes the magnetization to the remaining Mn content. With $S = 5/2$ Mn and assuming $g=2$, $M(x) = \frac{xSg}{1-x}$ becomes $M(x) = \frac{5x}{1-x}$. This equation is then used to saturation magnetization as a function of composition in figure S15.

\section{Supplemantry Note 3: SHG polarization depdence}

Under normal-incidence and parallel-polarization geometry, electric quadrupolar SHG takes the form $I(2\omega, \phi) \propto ((\chi_{xyzy}+\chi_{yxzy}+\chi_{yyzz})\cos^2\phi\sin\phi+\chi_{xxzx}\sin^3\phi)^2$, where $\phi$ corresponds to the angle between the incoming/outgoing beams and the crystallographic a-axis and $\chi_{ijkl}^{EQ}$ are elements from the electric quadrupole susceptibility tensor from point group 2/m.\supercite{Chu2020}

Space groups \textit{C}2 and \textit{P}2$_1$ belong to the point group 2. The elements of the electric dipole nonlinear susceptibility tensor, $\chi_{ijk}^{ED}$, for point group 2 are as follows for the 2-fold axis along y:\supercite{Boyd2008}

\begin{center}    
    $\begin{pmatrix}
        \biggr(\begin{smallmatrix}
        0\\
        \chi_{xyx}\\
        0
        \end{smallmatrix}\biggr)
        &  \biggr(\begin{smallmatrix}
        \chi_{xxy}\\
        0\\
        \chi_{xzy}
        \end{smallmatrix}\biggr) & \biggr(\begin{smallmatrix}
        0\\
        \chi_{xyz}\\
        0
        \end{smallmatrix}\biggr)\\
        \biggr(\begin{smallmatrix}
        \chi_{yxx}\\
        0\\
        \chi_{yzx}
        \end{smallmatrix}\biggr)
        & \biggr(\begin{smallmatrix}
        0\\
        \chi_{yyy}\\
        0
        \end{smallmatrix}\biggr) & \biggr(\begin{smallmatrix}
        \chi_{yxz}\\
        0\\
        \chi_{yzz}
        \end{smallmatrix}\biggr)\\
        \biggr(\begin{smallmatrix}
        0\\
        \chi_{zyx}\\
        0
        \end{smallmatrix}\biggr)
        & \biggr(\begin{smallmatrix}
        \chi_{zxy}\\
        0\\
        \chi_{zzy}
        \end{smallmatrix}\biggr) & \biggr(\begin{smallmatrix}
        0\\
        \chi_{zyz}\\
        0
        \end{smallmatrix}\biggr)\\
    \end{pmatrix}$
\end{center}

In normal incidence, parallel polarization geometry ($E_z = 0, E_x = \cos\phi, E_y = \sin\phi$) and using $E_i(2\omega) =\chi_{ijk}^{ED}E_j(\omega)E_k(\omega)$ expected SHG intensity observed at the detector is calculated as follows, where $\phi$ corresponds the angle between the incoming/outgoing beams and the crystallographic a-axis:$$E_x(2\omega, \phi) = (\chi_{xxy}\cos\phi\sin\phi + \chi_{xyx}\sin\phi\cos\phi)\cos\phi$$
$$E_y(2\omega, \phi) = (\chi_{yxx}\cos^2\phi + \chi_{xyy}\sin^2\phi)\sin\phi$$
$$I(2\omega, \phi) \propto (E_x(2\omega, \phi)+E_y(2\omega, \phi))^2$$
Therefore, the intensity of SHG signal takes the form
$$I(2\omega, \phi) \propto ((\chi_{xxy}+\chi_{xyx}+\chi_{yxx})\cos^2\phi\sin\phi+\chi_{xyy}\sin^3\phi)^2$$
The electric quadrupole SHG and electric dipole SHG take on the same functional form and accordingly equivalently fit the polarization data.

\pagebreak
\section{Supplementary Table 1: Raman mode energies}

Table S1: Energy of select MnPS\textsubscript{3} Raman modes from bulk crystals before and after intercalation

\begin{table}[h]
    \centering
    \begin{tabular}{ccc}
         Mode & Pristine MnPS\textsubscript{3} (cm$^{-1}$) & \textit{S} Intercalated (cm$^{-1}$) \\
         P1& 117 & 112\\
         P2&  156& 155\\
         P3&  224& 214\\
         P4&  246& 237\\
         P5&  275& 274\\
         P6&  386& 386\\
         P7&  568& 559\\
         P8&  582& 577\\
    \end{tabular}
    \label{tab:my_label}
\end{table}

\pagebreak

\section{Supplemental Figures}

\begin{figure*}[tbhp]
    \centerline{\includegraphics[]{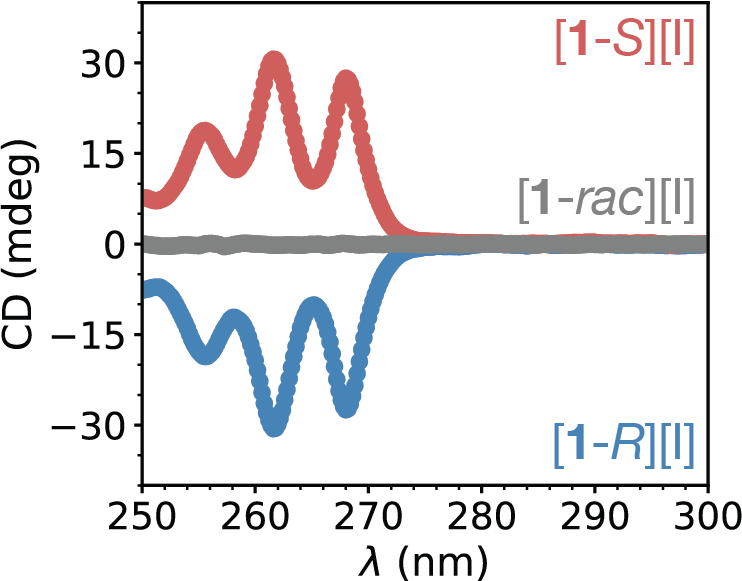}}
    \caption{\textbf{Circular dichroism spectra of intercalant molecules.}
    }\label{fig:S1}
\end{figure*}

\begin{figure*}[tbhp]
    \centerline{\includegraphics[]{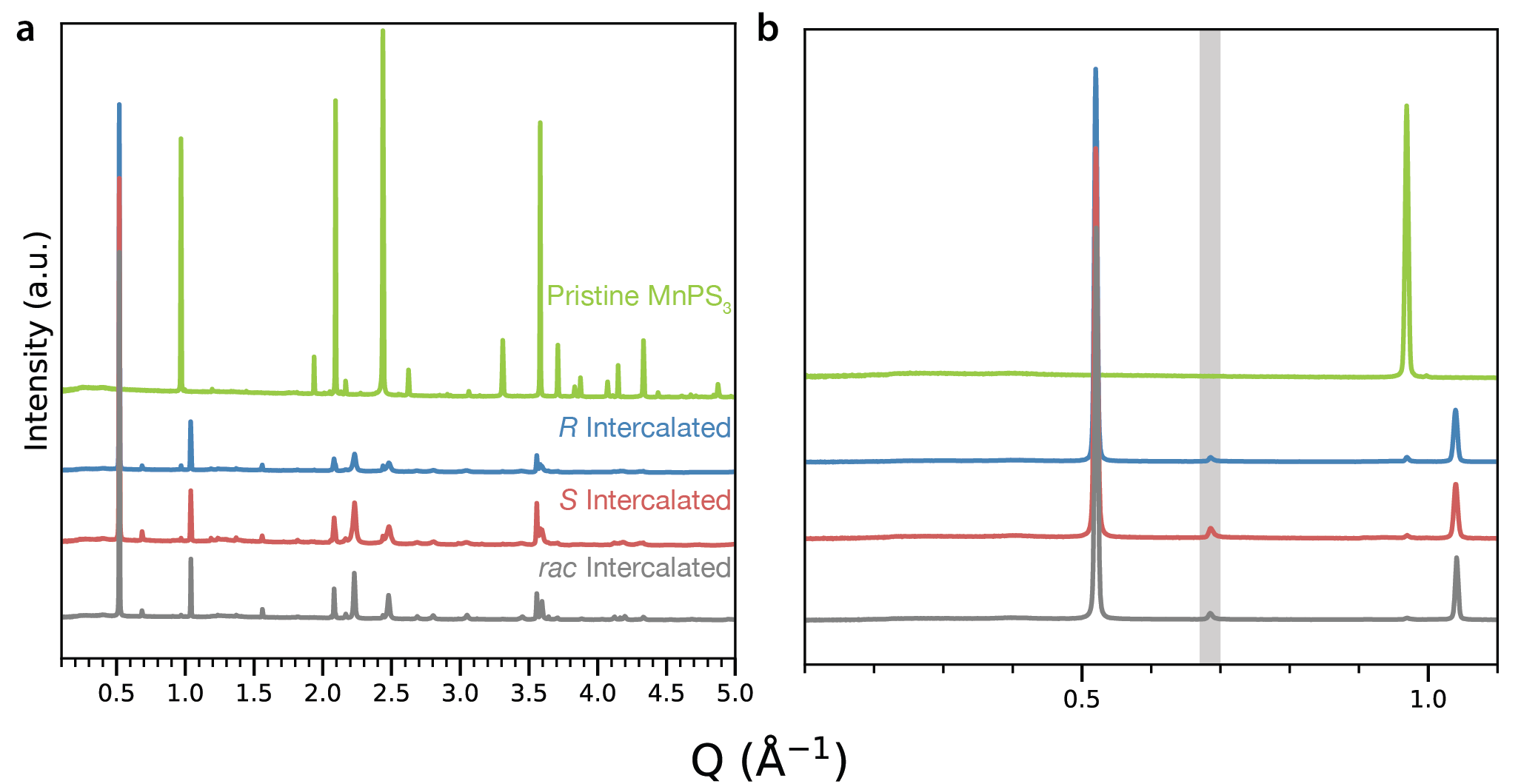}}
    \caption{
    \textbf{High–resolution, synchrotron PXRD patterns of powder intercalates.} \textbf{(a)} Full patterns \textbf{(b)} and low angle regions. All patterns are normalized to the maximum intensity reflection for clarity. The highlighted reflection appears only in intercalates and corresponds to a 9.16 Å spacing. This reflection has been attributed both to a superlattice and an impurity in previous literature\supercite{Evans1995, Iton2024}. Small amounts of unintercalated phase are visible in the low–angle region.}\label{fig:S2}
\end{figure*}

\begin{figure*}[tbhp]
    \centerline{\includegraphics[width=160mm]{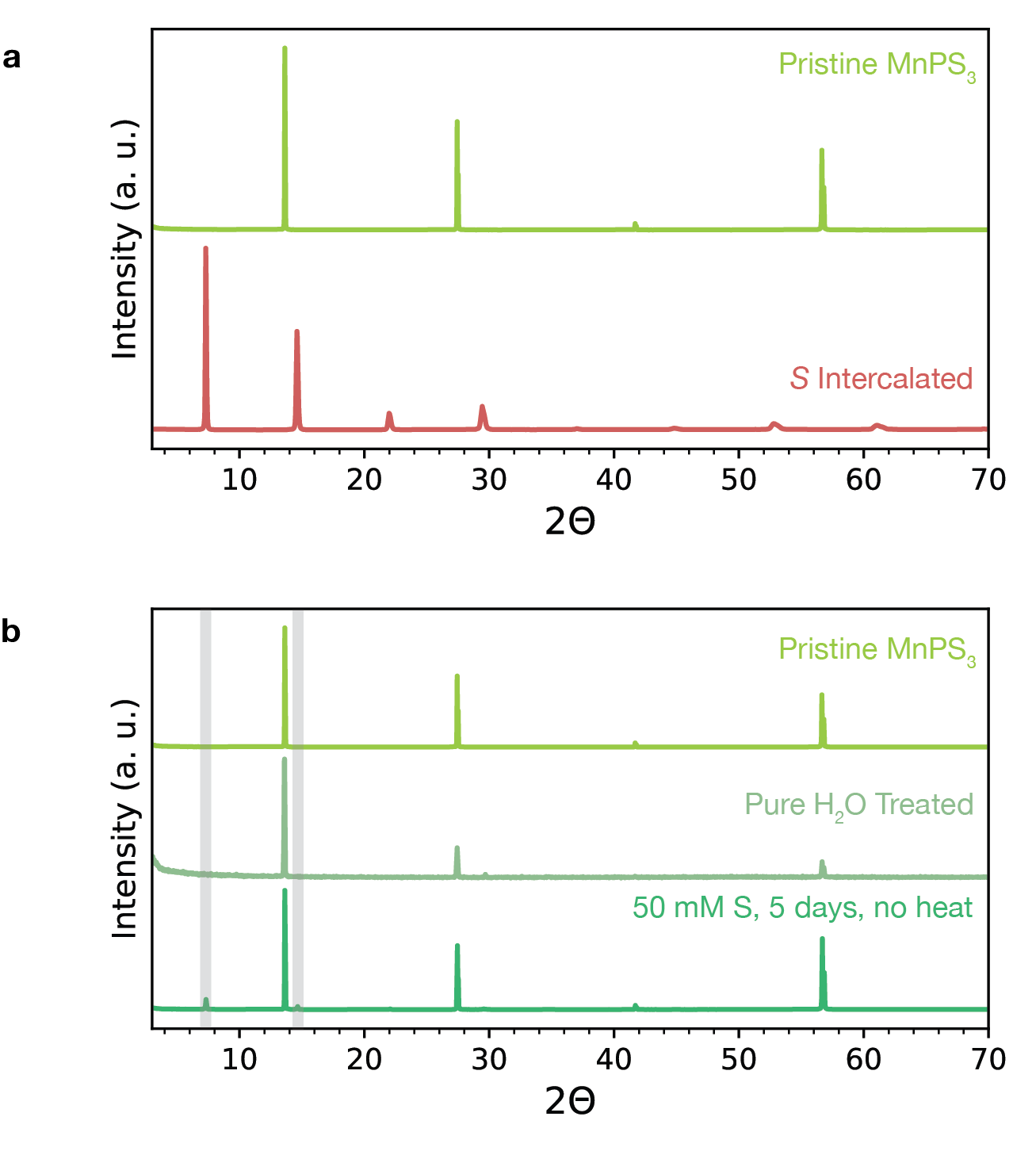}}
    \caption{
    \textbf{Diffraction patterns of bulk crystal intercalates.} \textbf{(a)} Diffraction demonstrating the complete intercalation of MnPS\textsubscript{3} via complete replacement of the (00l) family of reflections. \textbf{(b)} Diffraction pattern showing the necessity of both heat and chiral cations to change the interlayer spacing of MnPS\textsubscript{3}. Highlighted reflections are characteristic of a minor intercalated phase. Intensities of all patterns are normalized to the pristine (001) reflection for comparison.
    }\label{fig:S3}
\end{figure*}

\begin{figure*}[tbhp]
    \centerline{\includegraphics[width=160mm]{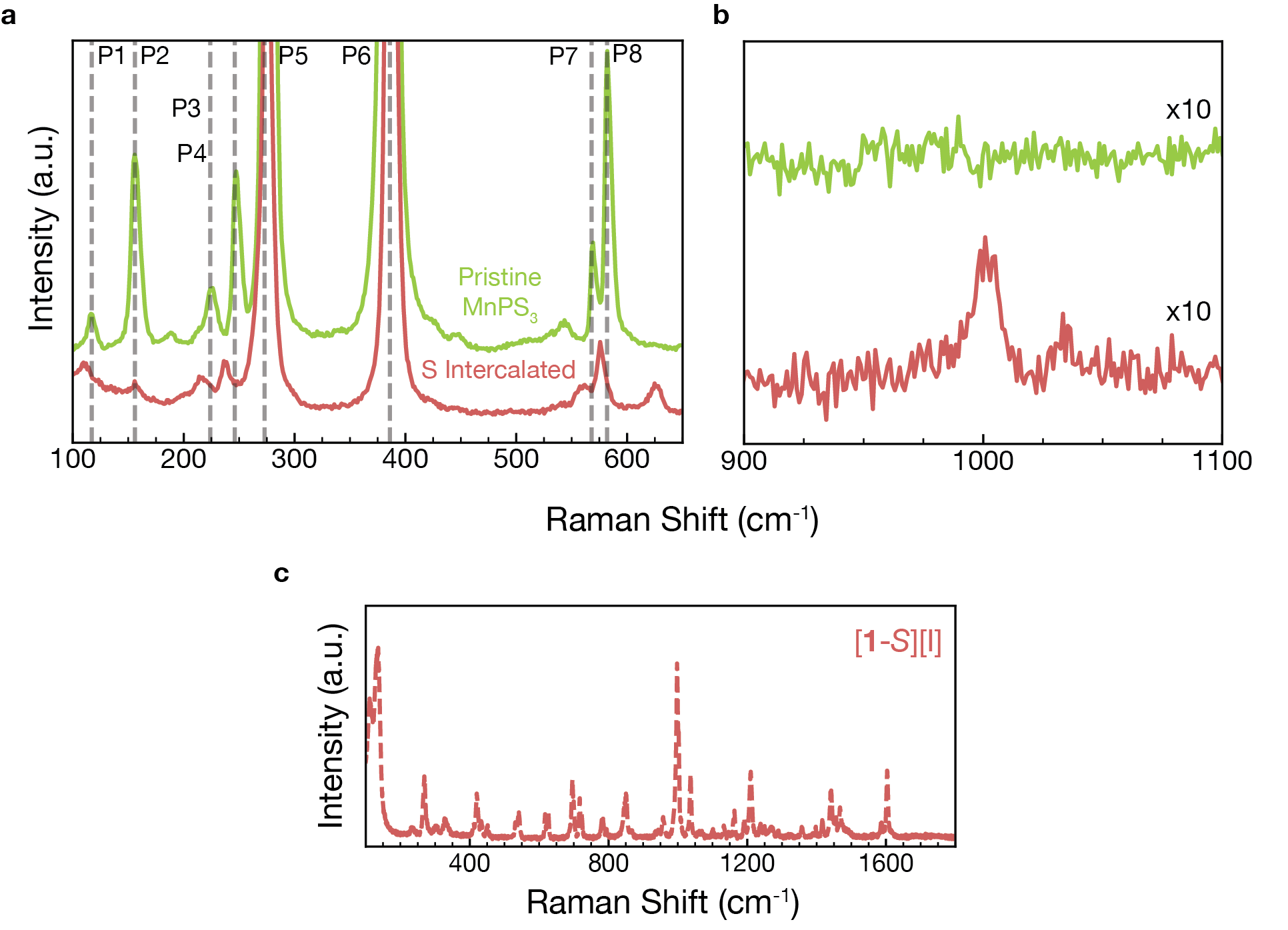}}
    \caption{ \textbf{Raman spectra of bulk crystals.}
    \textbf{(a)} Raman spectra of a bulk crystal of MnPS\textsubscript{3} and bulk crystal \textit{S} intercalate. Data is normalized to intensity of mode P6 for comparison. Energies of numbered modes are detailed in table S1. \textbf{(b)} Modes originating from the chiral intercalant molecule are weakly observed. \textbf{(c)} Raman of isolated \textbf{1}-\textit{S}.
    }\label{fig:S4}
\end{figure*}

\begin{figure*}[tbhp]
    \centerline{\includegraphics[width=160mm]{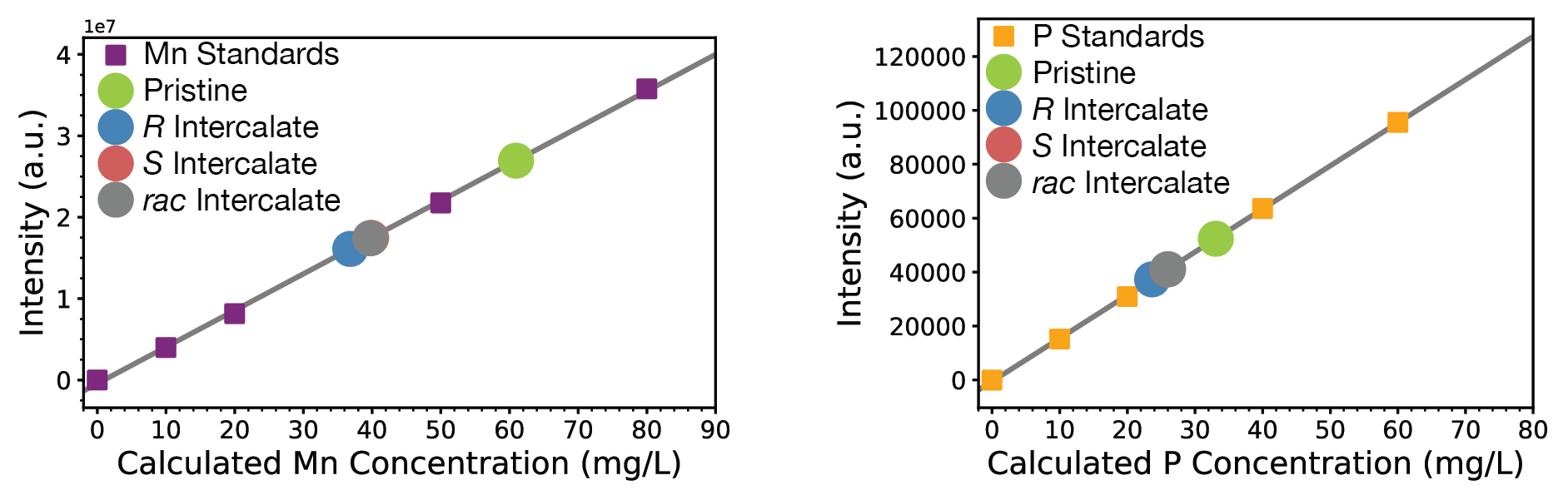}}
    \caption{\textbf{
    Representative Mn and P ICP data for compositional analysis.}
    }\label{fig:S5}
\end{figure*}

\begin{figure*}[tbhp]
    \centerline{\includegraphics[width=160mm]{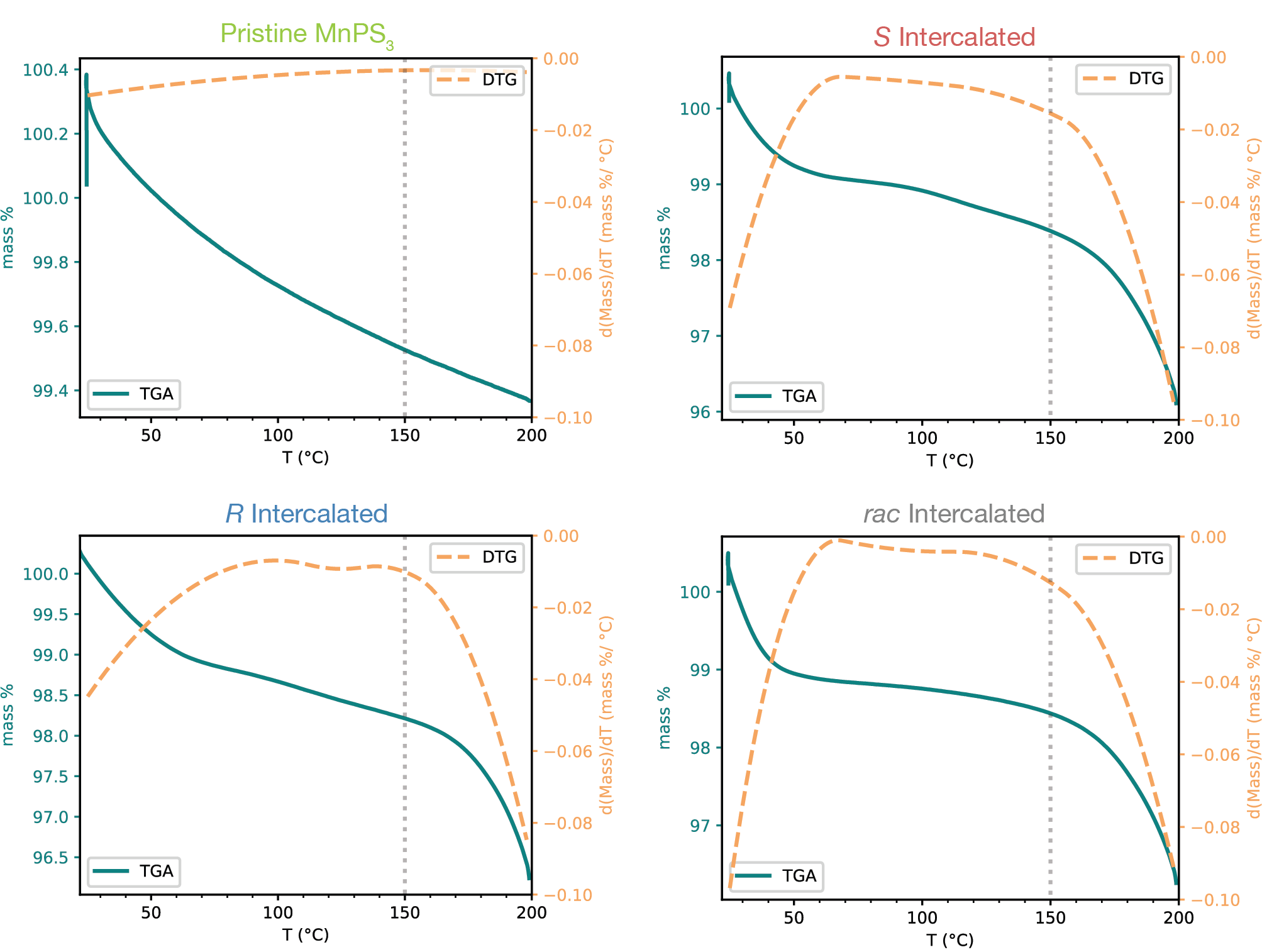}}
    \caption{
    \textbf{TGA and derivative thermogravimetry (DTG) of pristine MnPS\textsubscript{3} and intercalates.} Mass losses at 150 °C for pristine, \textit{R}, \textit{S}, and \textit{rac} are  0.47 \%, 1.79 \%, 1.61 \%, 1.56 \% respectively.
    }\label{fig:S6}
\end{figure*}

\begin{figure*}[tbhp]
    \centerline{\includegraphics[width=160mm]{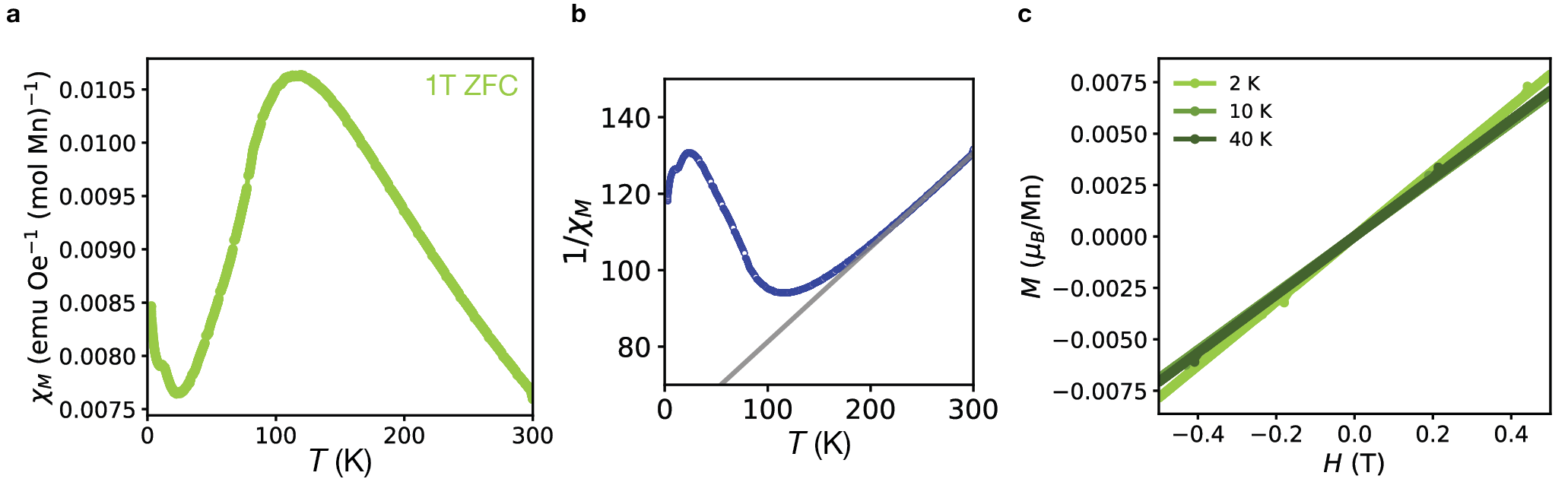}}
    \caption{
    \textbf{Magnetometry data for pristine MnPS\textsubscript{3}.} \textbf{(a)} Low overall moment, a broad cusp in $\chi$ vs T, \textbf{(b)} and a negative Curies–Weiss temperature indicate the material is an antiferromagnet. Data was collected with an applied field of 1 T. \textbf{(c)} The MnPS\textsubscript{3} shows no hysteresis and minimal uncompensated moment in \textit{M} vs \textit{H} sweeps at temperatures relevant to ferrimagnetic ordering of the intercalated samples.
    }\label{fig:S7}
\end{figure*}

\begin{figure*}[tbhp]
    \centerline{\includegraphics[width=160mm]{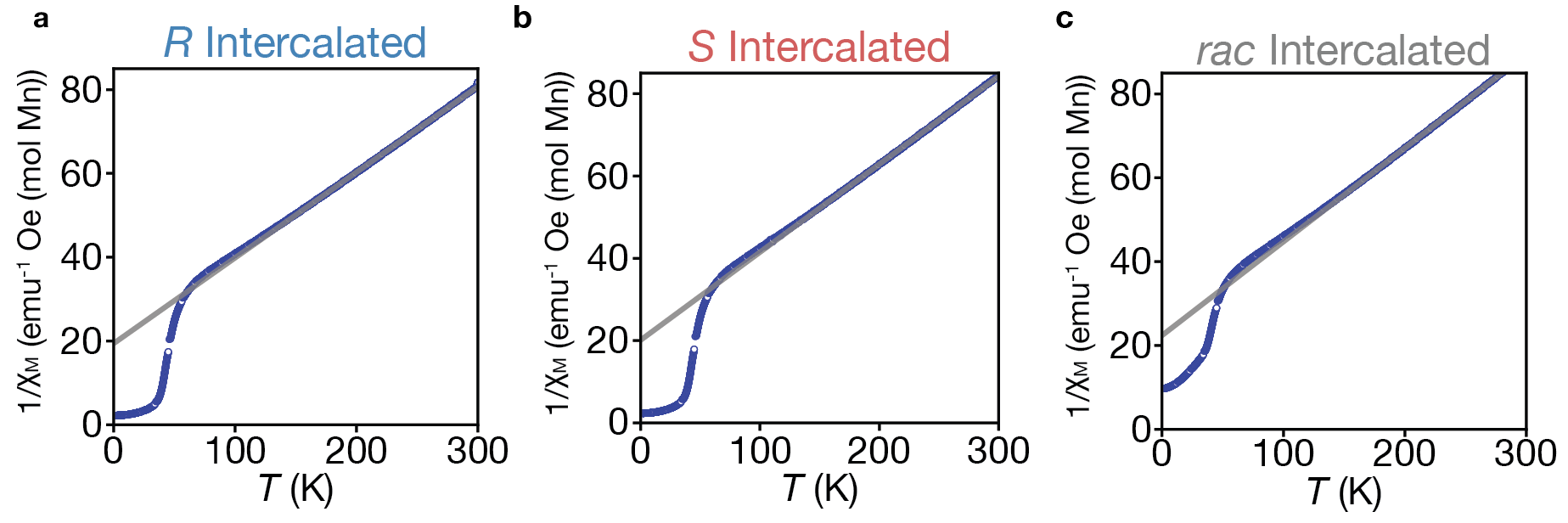}}
    \caption{
    \textbf{Curie--Weiss fits to samples from Figure 2a.} Data was collected with 1 T applied field.
    }\label{fig:E1}
\end{figure*}

\begin{figure*}[tbhp]
    \centerline{\includegraphics[]{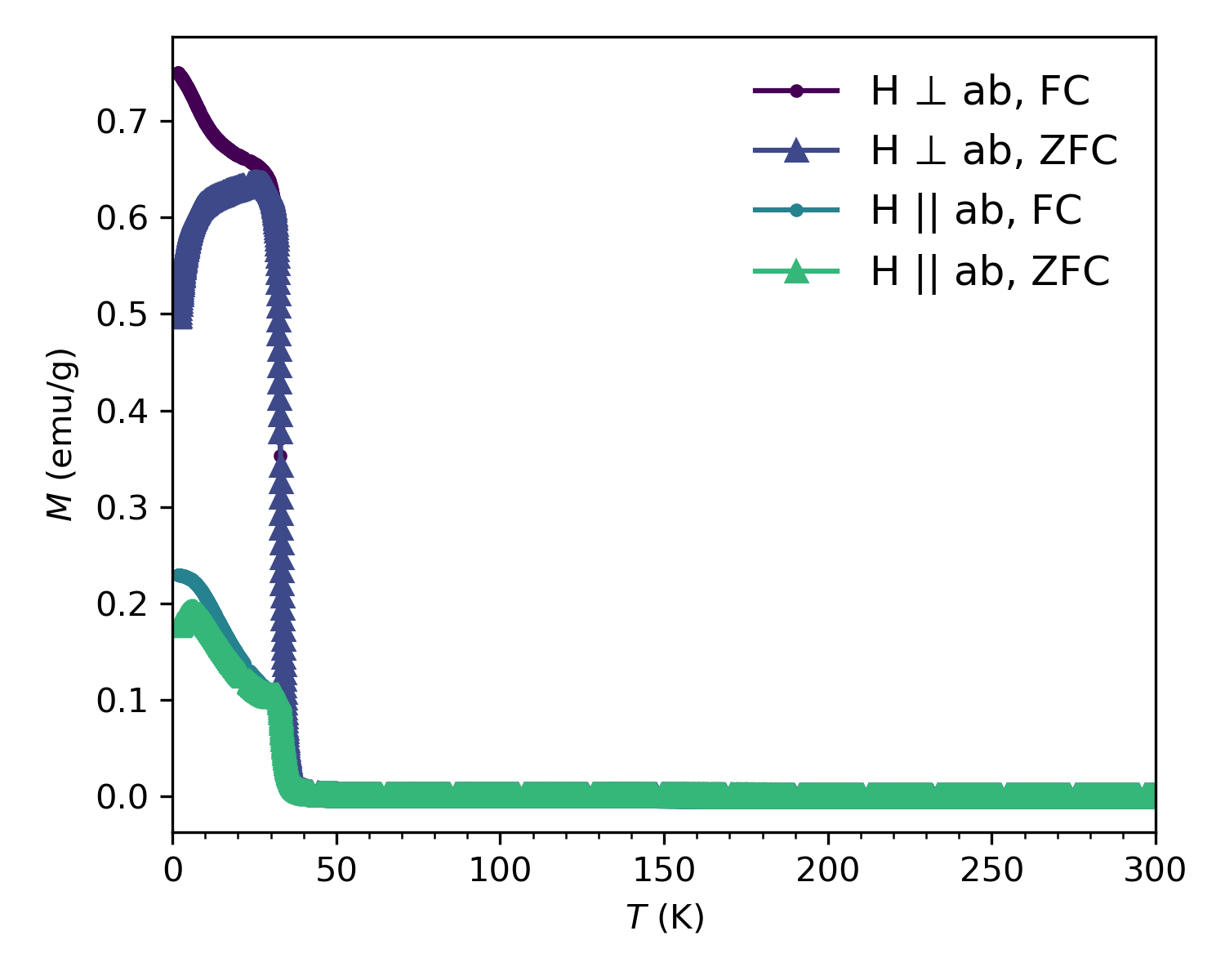}}
    \caption{
    \textbf{Magnetometry from a collection of \textit{S} intercalated MnPS\textsubscript{3} bulk crystals.} Samples are all measured with a 100 Oe applied field. Magnetization is reported as per gram of sample (not per gram of Mn). Samples were measured with both an out of plane ($H$ $\perp$ ab) and in–plane ($H$ $||$ ab) field.
    }\label{fig:S8}
\end{figure*}

\begin{figure*}[tbhp]
    \centerline{\includegraphics[width=160mm]{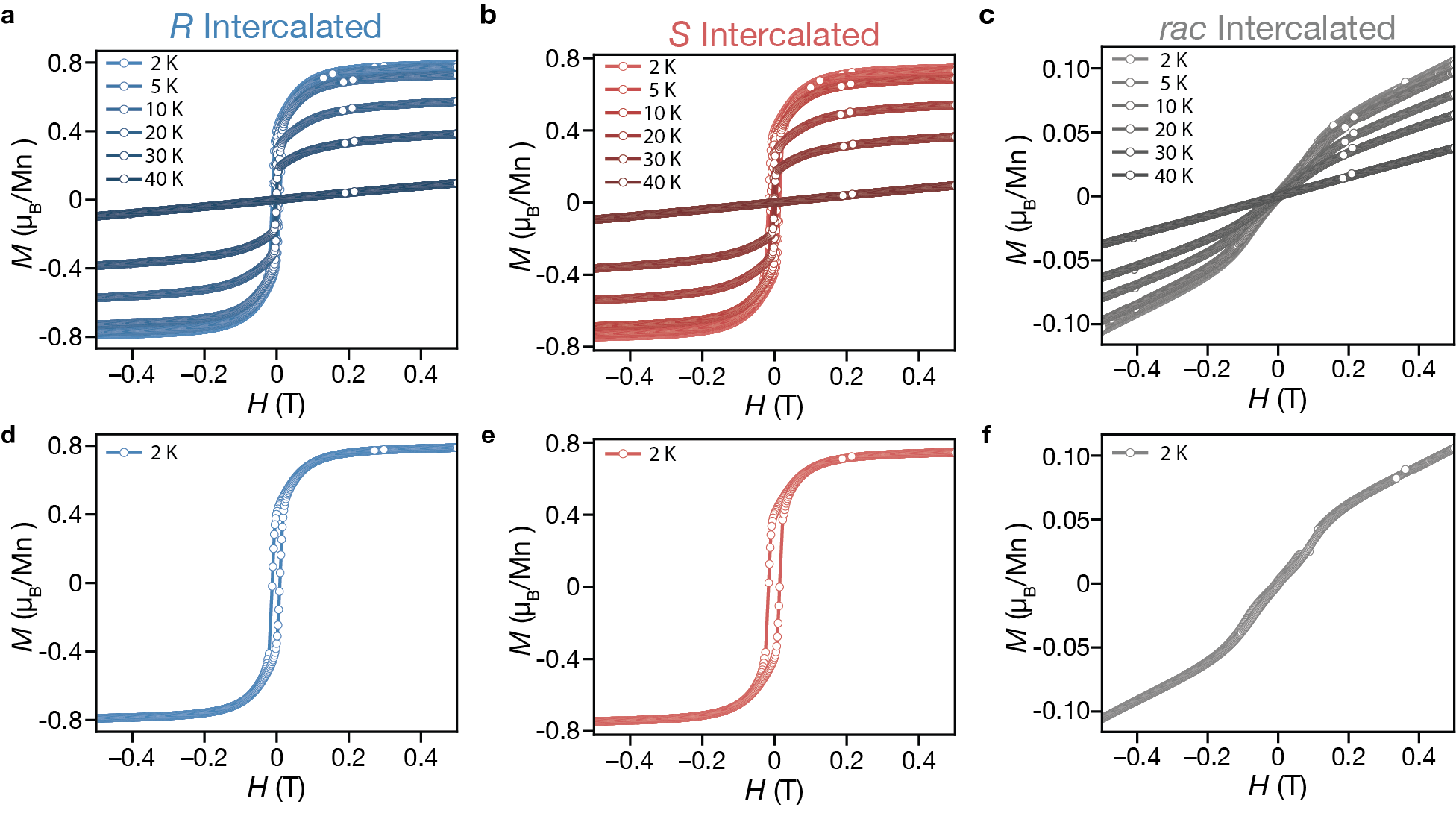}}
    \caption{
    \textbf{Temperature dependence of \textit{M} vs. \textit{H} curves of MnPS\textsubscript{3} intercalates.} \textbf{(a-c)} All data and \textbf{(d-f)} data collected at 2 K to show the materials' small, but finite hysteresis.
}\label{fig:E1}
\end{figure*}

\begin{figure*}[tbhp]
    \centerline{\includegraphics[width=160mm]{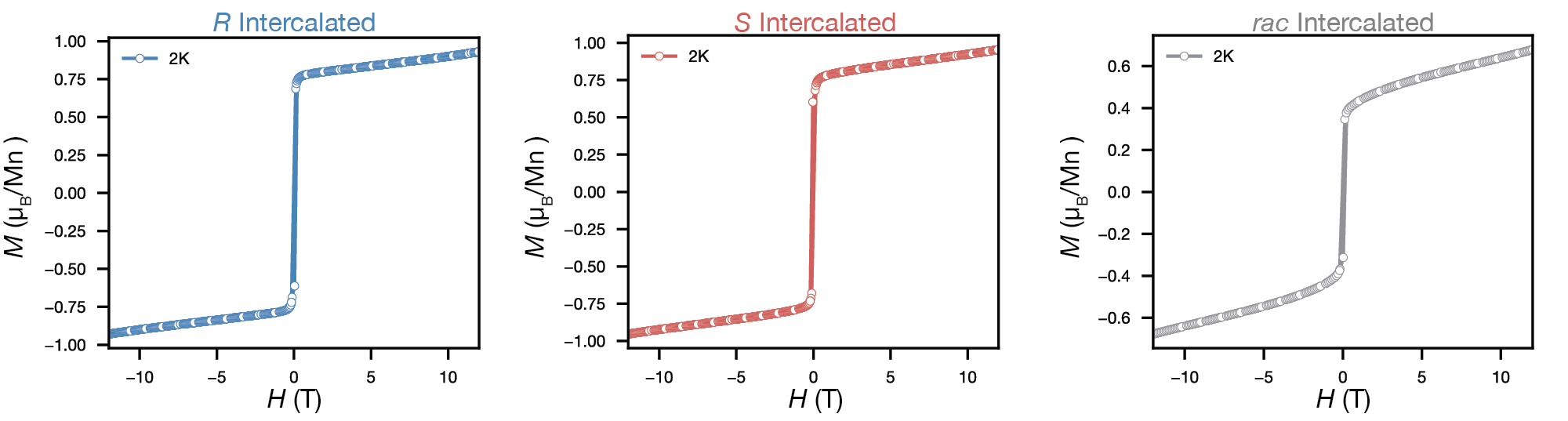}}
    \caption{
    \textbf{\textit{M} vs. \textit{H} curves of MnPS\textsubscript{3} intercalates measured to ±12 T at 2 K.} 
    }\label{fig:S9}
\end{figure*}

\begin{figure*}[tbhp]
    \centerline{\includegraphics[width=160mm]{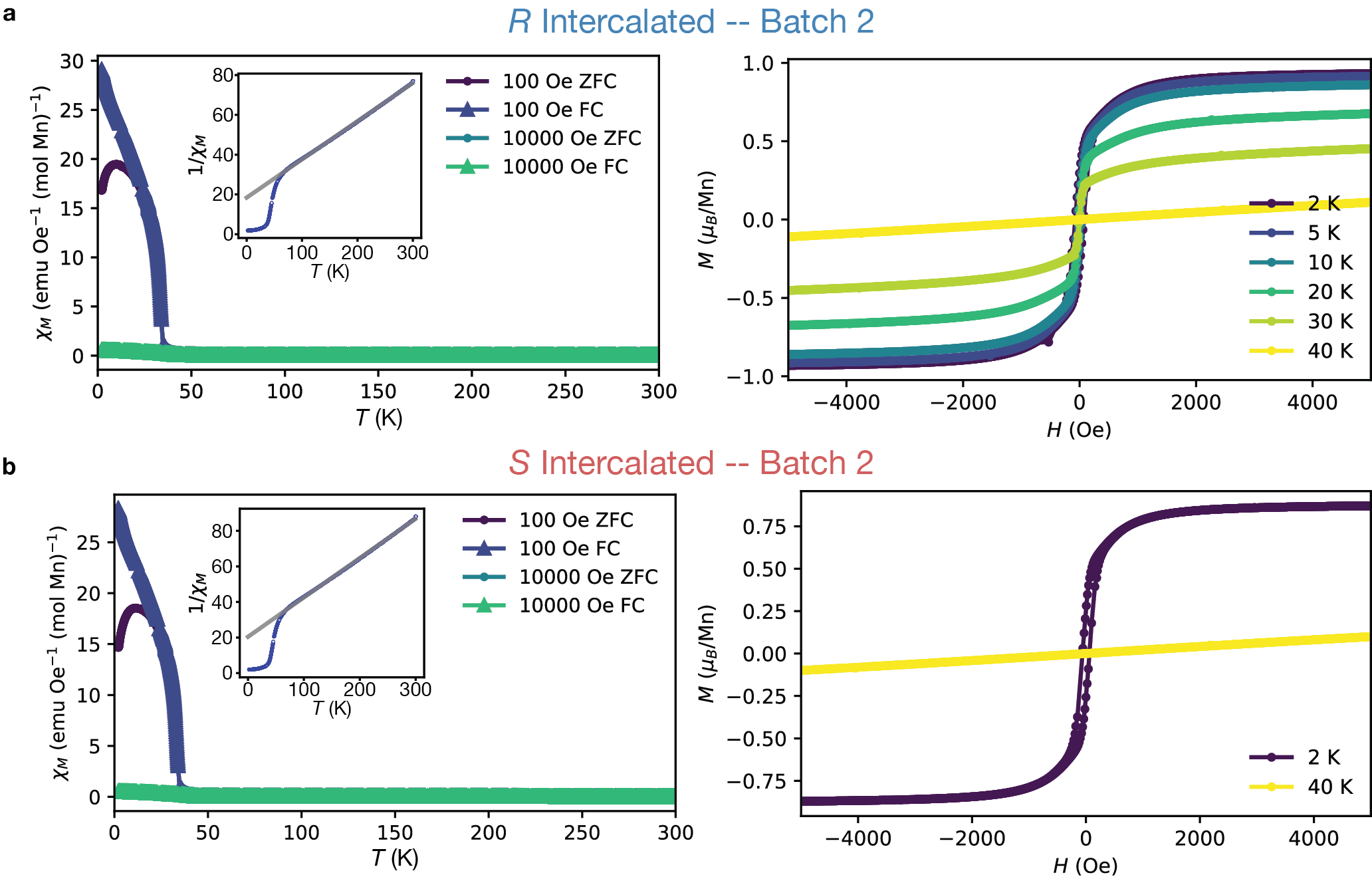}}
    \caption{
    \textbf{Additional magnetometry of enantiopure intercalates.} All FC measurements for this data set were done via field cooling in a 1 T field. Curie Weiss fitting was conducted on data collected at 1 T applied field.
    }\label{fig:S10} 
\end{figure*}

\begin{figure*}[tbhp]
    \centerline{\includegraphics[width=160mm]{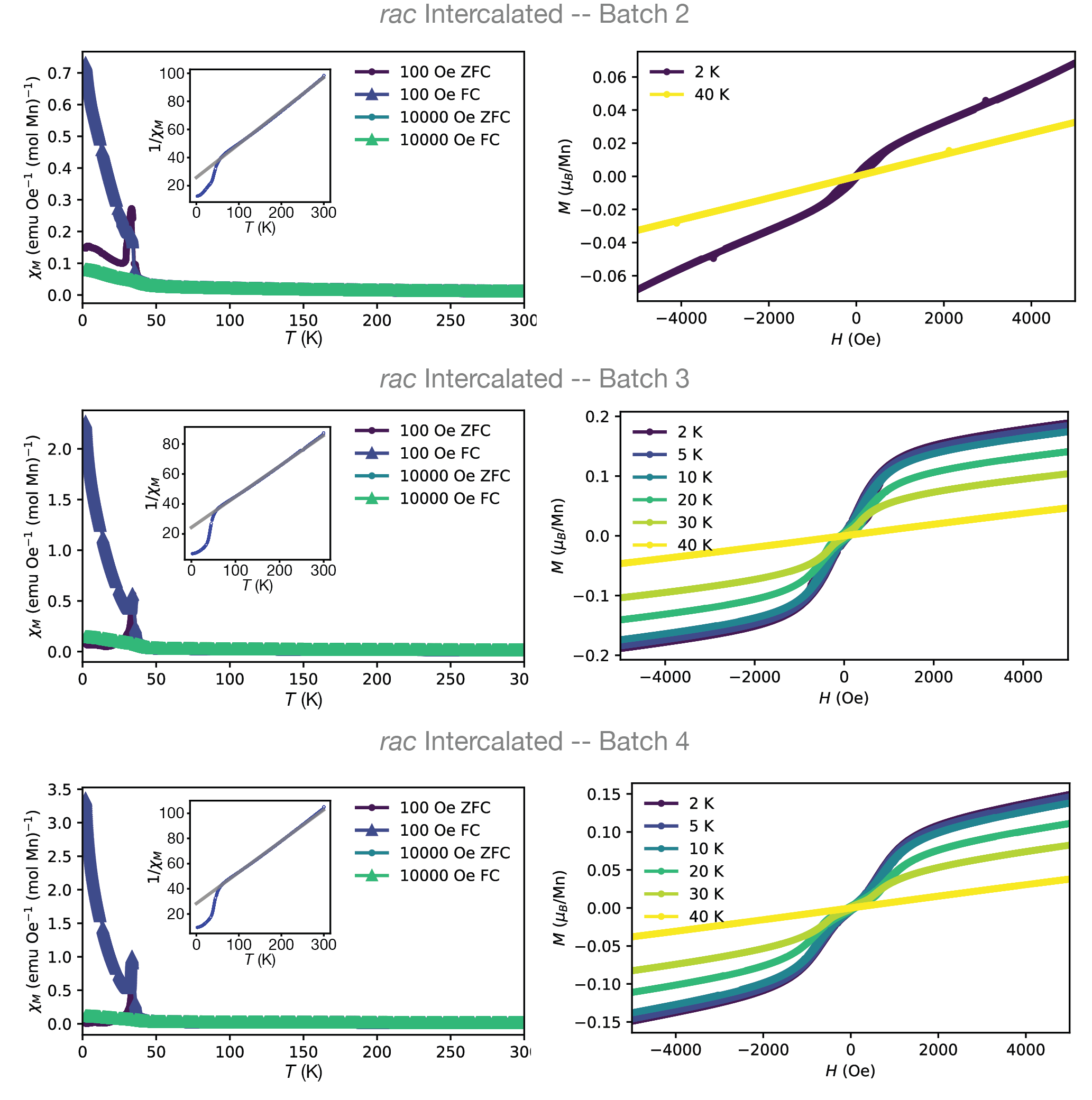}}
    \caption{
    \textbf{Additional magnetometry of racemic intercalates.} All FC measurements for this data set were done via field cooling in a 1 T field. Curie Weiss fitting was conducted on data collected at 1 T applied field.
    }\label{fig:S11}
\end{figure*}

\begin{figure*}[tbhp]
    \centerline{\includegraphics[]{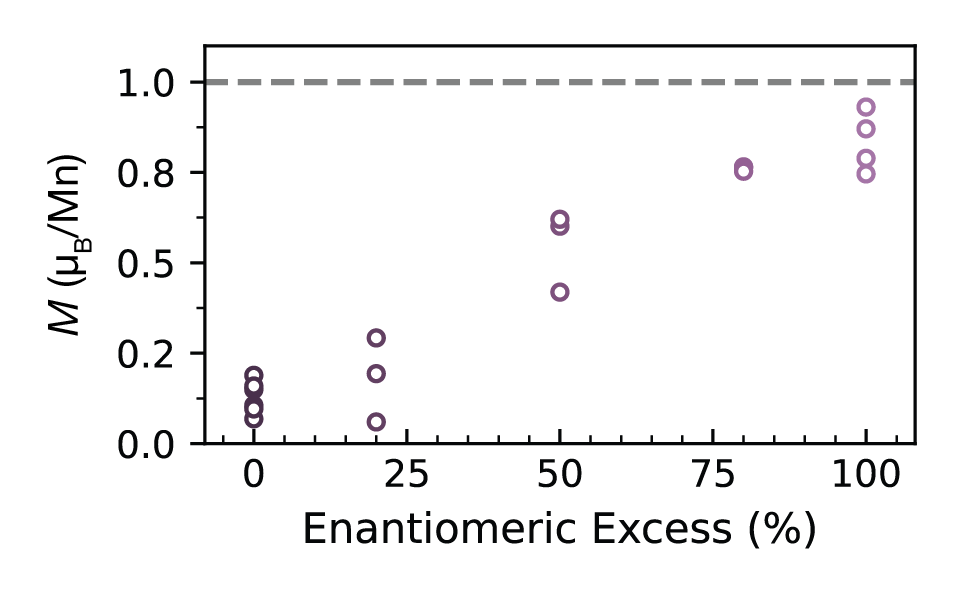}}
    \caption{
    \textbf{Maximum magnetic moment observed from intercalates of different enantiopurity at 2 K and 0.5 T.} The dashed line is the idealized max magnetic moment from a 1/6 replacement of Mn in MnPS\textsubscript{3}.
    }\label{fig:S12}
\end{figure*}

\begin{figure*}[tbhp]
    \centerline{\includegraphics[]{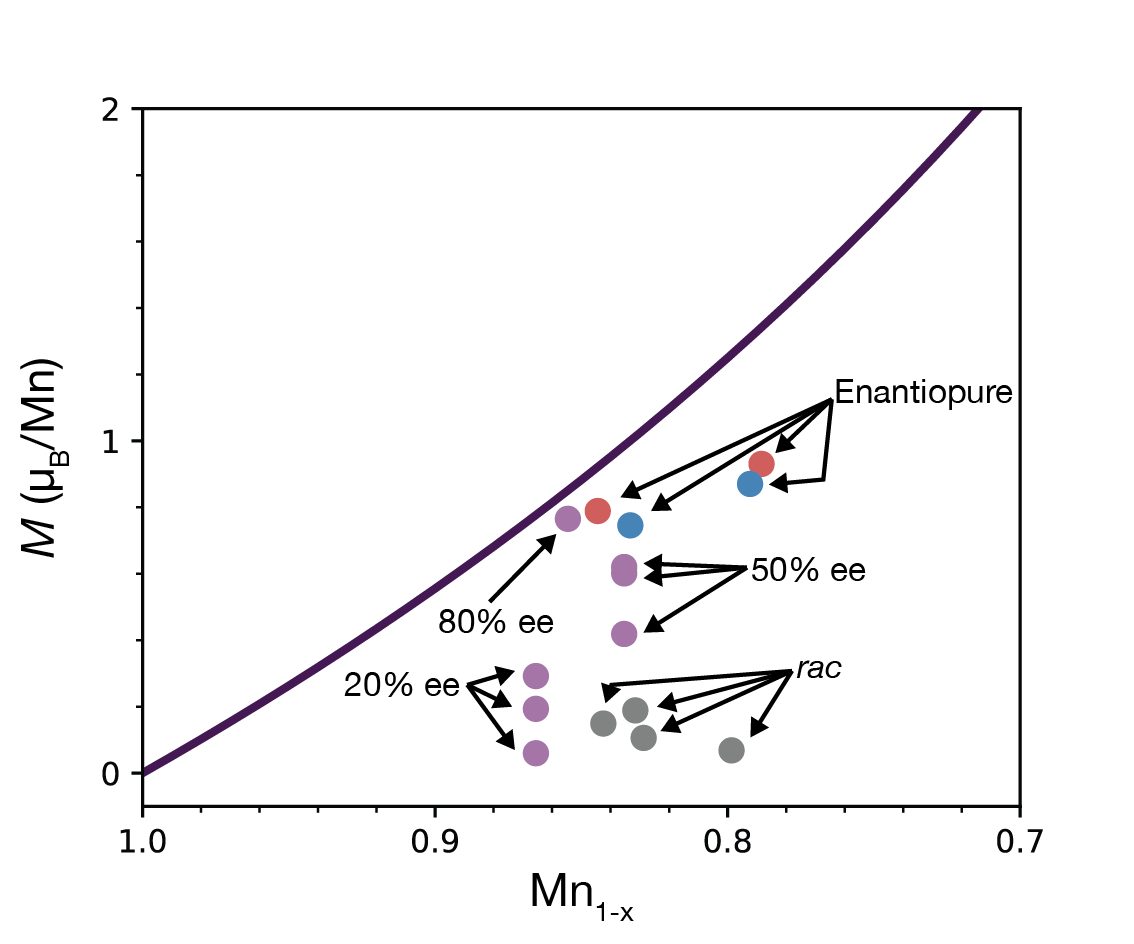}}
    \caption{
    \textbf{Measured magnetization at 0.5 T and 2 K of unaged, different chirality intercalates as a function of remaining Mn amount after intercalation.} The purple curve is the theoretical saturation magnetization based on composition as described in supplementary note 2. There is no clear trend between intercalate Mn composition and observed magnetization. 
    }\label{fig:S13}
\end{figure*}

\begin{figure*}[tbhp]
    \centerline{\includegraphics[]{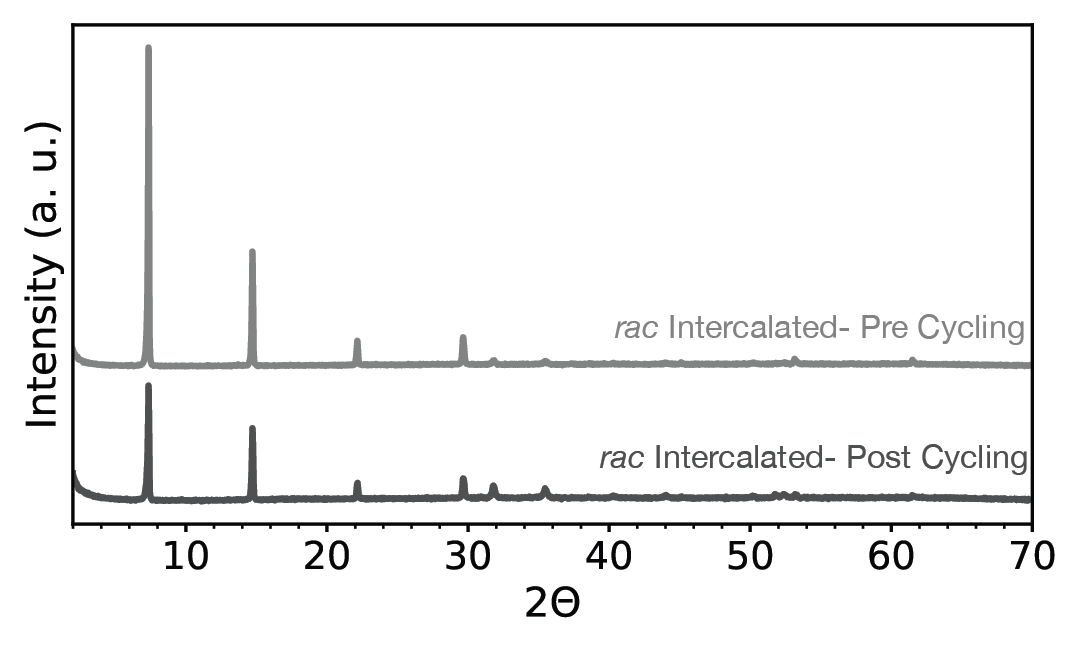}}
    \caption{
    \textbf{PXRD data of a \textit{rac} intercalated sample before and after thermal cylcing measurements in the PPMS.} No new reflections appear, indicating the overall host lattice structure is unchanged by cycling.
    }\label{fig:S14}
\end{figure*}

\begin{figure*}[tbhp]
    \centerline{\includegraphics[width=160mm]{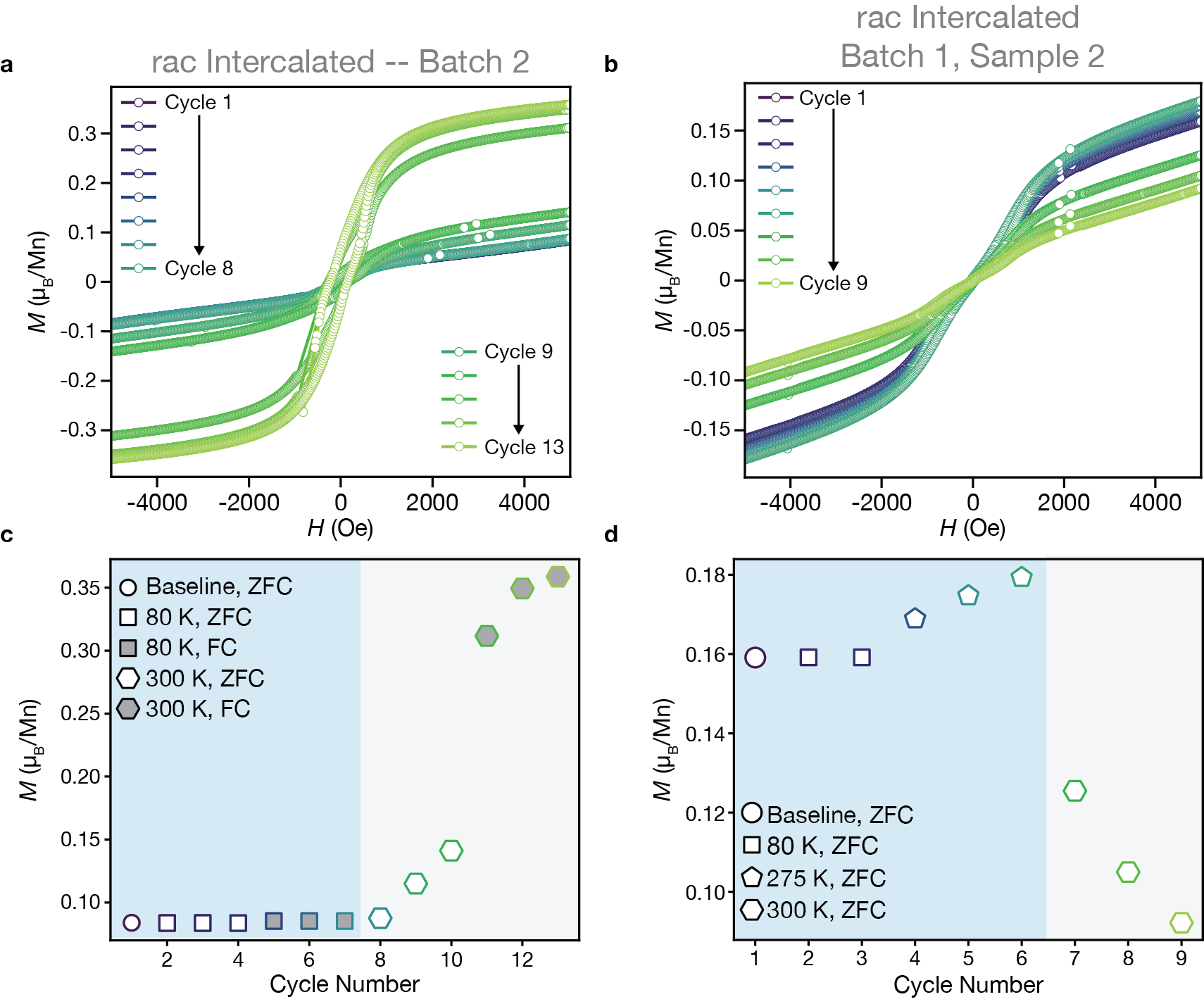}}
    \caption{
    \textbf{Additional thermal cycling data for \textit{rac} intercalated MnPS\textsubscript{3}}. \textbf{(a-b)} $M$ vs. $H$  curves and \textbf{(c-d)} corresponding maximum magnetization values at 0.5 T. Blue shading indicates thermal cycles conducted under 300 K. FC indicates the sample was cooled in the presence of a 12 T field. Field cooling is not necessary to induce magnetization evolution. Isothermal dwell times for the left and right plots were 30 seconds and 30 minutes, respectively. Panels \textbf{(a)} and \textbf{(b)} are downsampled by 2 for ease of interpretation.
    }\label{fig:E1}
\end{figure*}

\begin{figure*}[tbhp]
    \centerline{\includegraphics[]{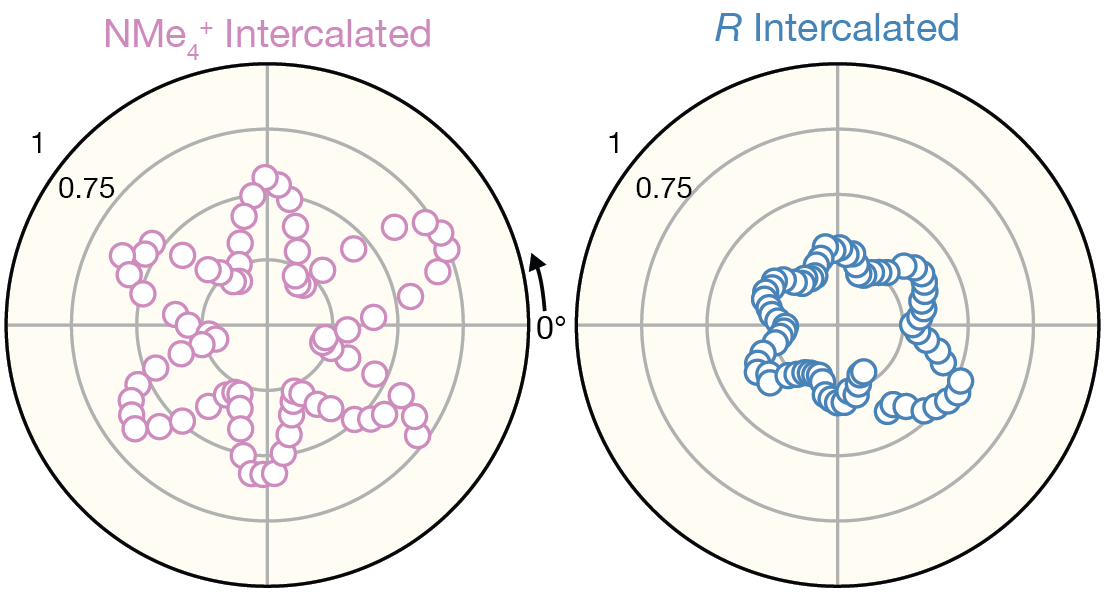}}
    \caption{
    \textbf{SHG response from a NMe\textsubscript{4}\textsuperscript{+} intercalated bulk flake and a \textit{R} intercalated flake.} Both samples are collected with the same experimental conditions.
    }\label{fig:S16}
\end{figure*}

\begin{figure*}[tbhp]
    \centerline{\includegraphics[]{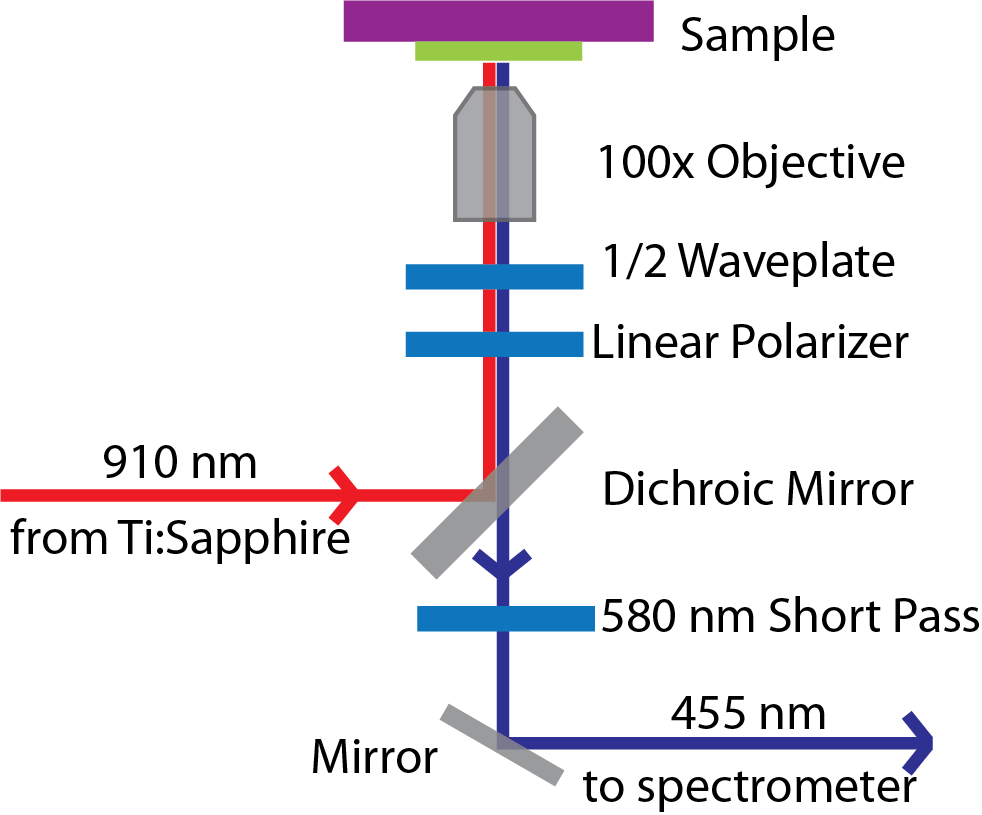}}
    \caption{
    \textbf{Schematic of the SHG measurement geometry.} 
    }\label{fig:S17}
\end{figure*}

\begin{figure*}[tbhp]
    \centerline{\includegraphics[width=160mm]{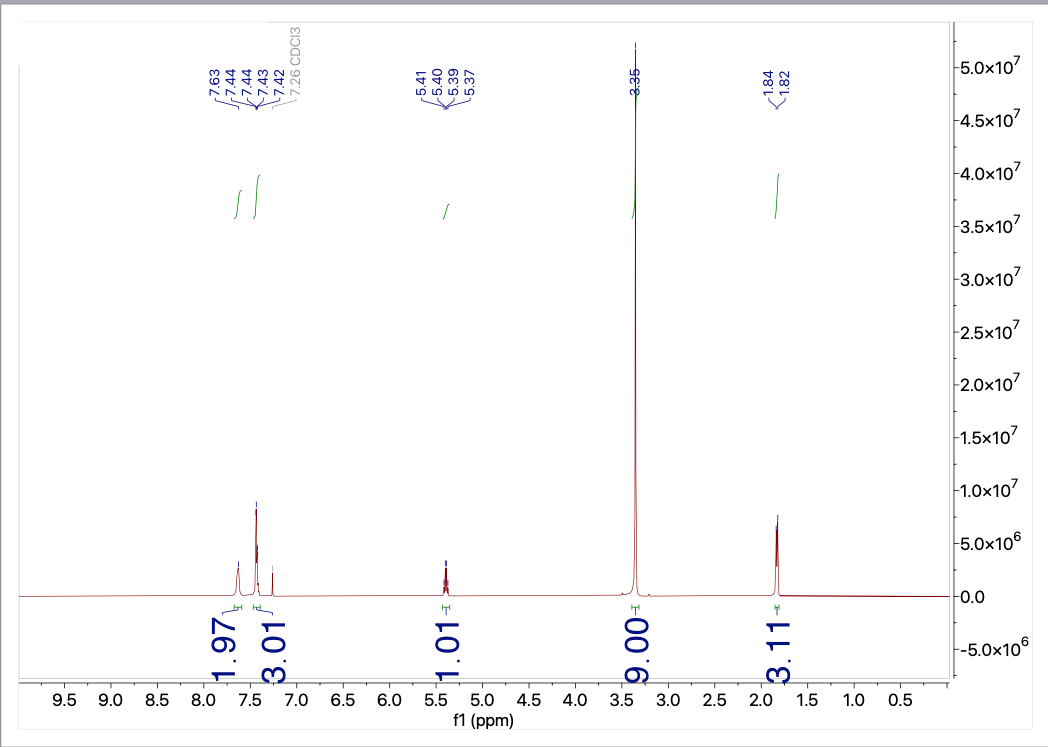}}
    \caption{
    \textbf{$^1$H NMR of [\textbf{1}-\textit{R}][I].} 
    }\label{fig:S18}
\end{figure*}

\begin{figure*}[tbhp]
    \centerline{\includegraphics[width=160mm]{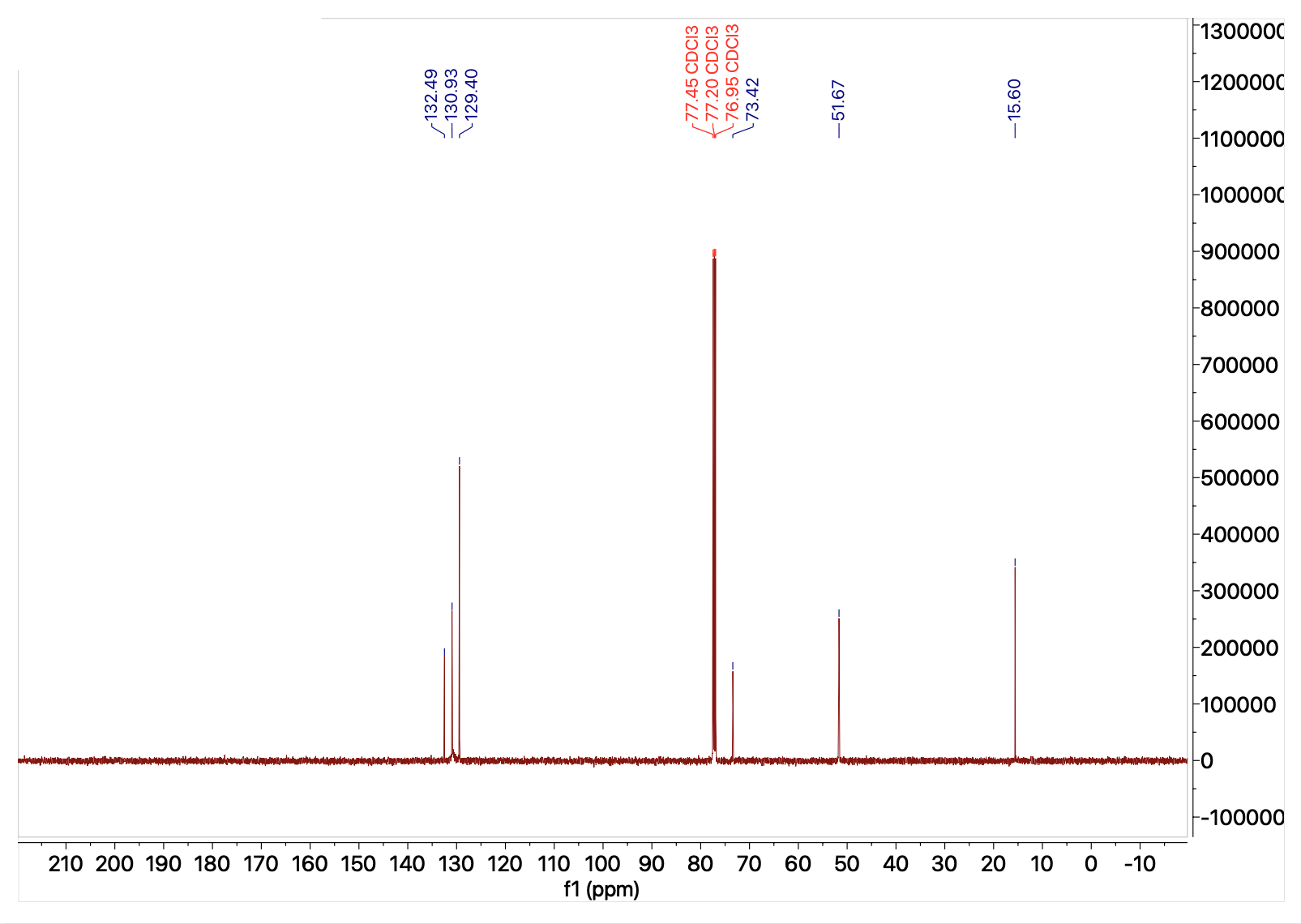}}
    \caption{
    \textbf{$^{13}$C NMR of [\textbf{1}-\textit{R}][I].} 
    }\label{fig:S19}
\end{figure*}

\begin{figure*}[tbhp]
    \centerline{\includegraphics[width=160mm]{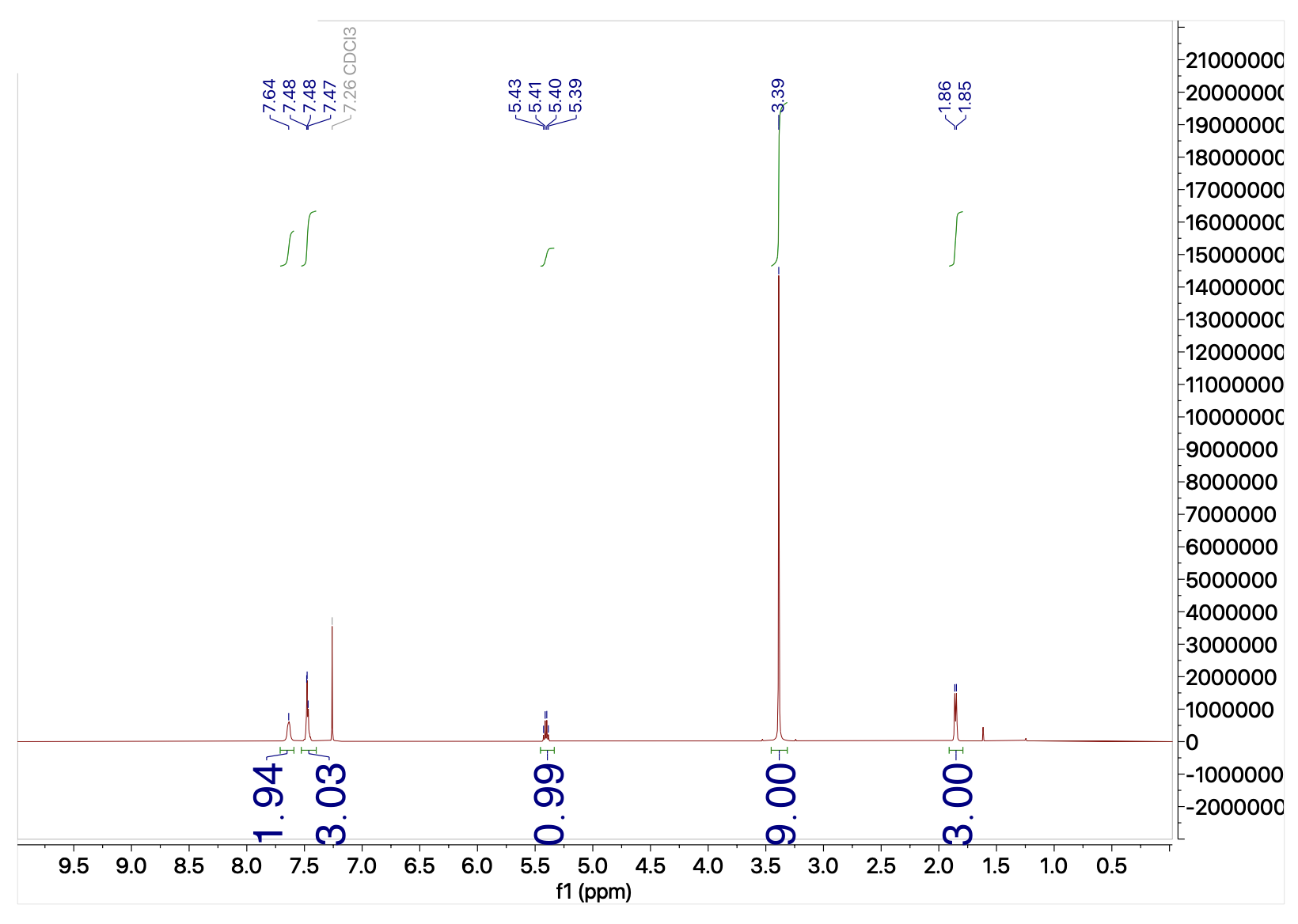}}
    \caption{
    \textbf{$^1$H NMR of [\textbf{1}-\textit{S}][I].} 
    }\label{fig:S20}
\end{figure*}

\begin{figure*}[tbhp]
    \centerline{\includegraphics[width=160mm]{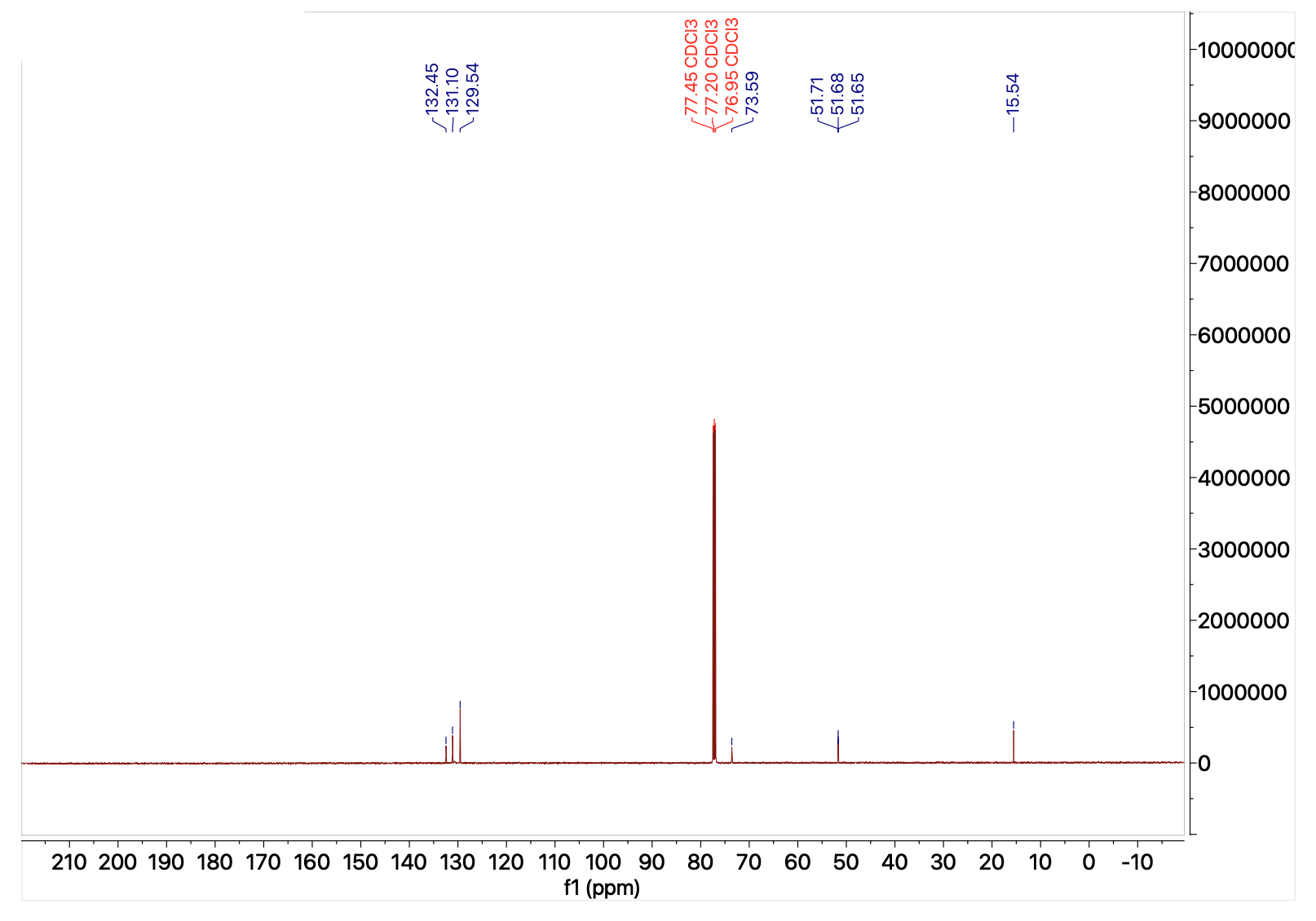}}
    \caption{
    \textbf{$^{13}$C NMR of [\textbf{1}-\textit{S}][I].} 
    }\label{fig:S21}
\end{figure*}

\begin{figure*}[tbhp]
    \centerline{\includegraphics[width=160mm]{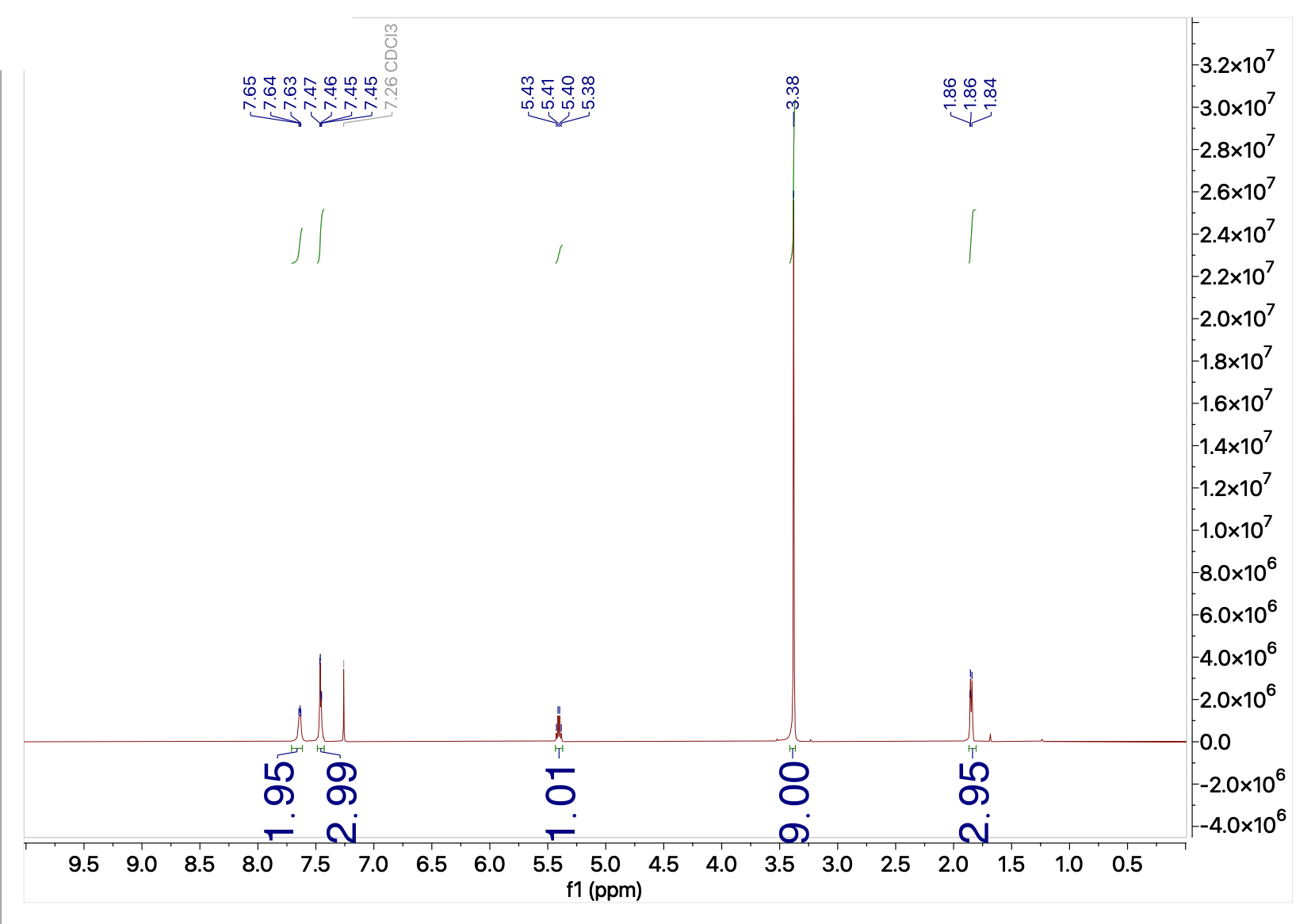}}
    \caption{
    \textbf{$^{1}$H NMR of [\textbf{1}-\textit{rac}][I].} 
    }\label{fig:S22}
\end{figure*}

\begin{figure*}[tbhp]
    \centerline{\includegraphics[width=160mm]{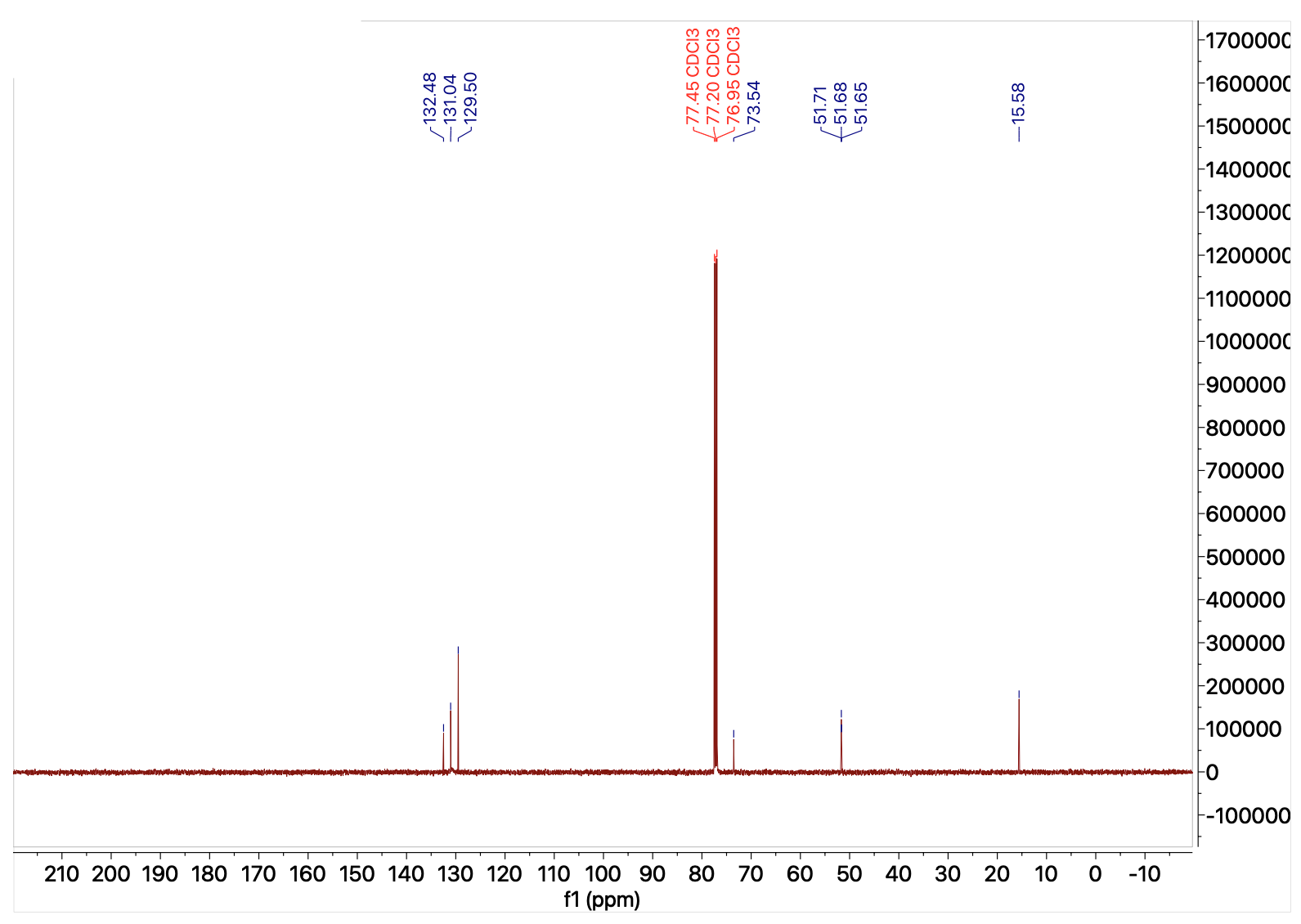}}
    \caption{
    \textbf{$^{13}$C NMR of [\textbf{1}-\textit{rac}][I].} 
    }\label{fig:S23}
\end{figure*}

\clearpage
\printbibliography